\documentclass[aps,prl,letterpaper,twocolumn,superscriptaddress,showpacs,amsmath,amssymb,nofootinbib,floatfix]{revtex4-1}
\usepackage{graphicx}  
\graphicspath{ {figures/} } 
\usepackage{bm}     
\usepackage{amssymb}   
\usepackage{amsmath}   
\usepackage{hyperref}
\usepackage{ulem}
\usepackage{natbib}

\begin{document}
\title{Improved neutron lifetime measurement with UCN$\tau$}
\newcommand{\indiana}{Department of Physics, Indiana University, Bloomington, IN, 47405, USA}
\newcommand{\ceem}{Center for Exploration of Energy and Matter, Indiana University, Bloomington, IN, 47405, USA}
\newcommand{\tunl}{Triangle Universities Nuclear Laboratory, Durham, NC 27708, USA}
\newcommand{\ornl}{Oak Ridge National Laboratory, Oak Ridge, TN 37831, USA}
\newcommand{\lanl}{Los Alamos National Laboratory, Los Alamos, NM, USA, 87545, USA}
\newcommand{\ncsu}{Department of Physics, North Carolina State University, Raleigh, NC 27695, USA}
\newcommand{\caltech}{Kellogg Radiation Laboratory, California Institute of Technology, Pasadena, CA 91125, USA}
\newcommand{\argonne}{Argonne National Laboratory, Lemont, IL 60439, USA}
\newcommand{\depauw}{DePauw University, Greencastle, IN 46135, USA}
\newcommand{\etsu}{East Tennessee State University, Johnson City, TN 37614, USA}
\newcommand{\ill}{Institut Laue-Langevin, CS 20156, 38042 Grenoble Cedex 9, France}
\newcommand{\jinr}{Joint Institute for Nuclear Research, 141980 Dubna, Russia}
\newcommand{\ttu}{Tennessee Technological University, Cookeville, TN 38505, USA}

\author{F.~M.~Gonzalez}\affiliation{\indiana}\affiliation{\ceem}\affiliation{\ornl}
\author{E.~M.~Fries}\affiliation{\caltech}
\author{C.~Cude-Woods}\affiliation{\ncsu}\affiliation{\tunl}
\author{T.~Bailey}\affiliation{\ncsu}\affiliation{\tunl}
\author{M.~Blatnik}\affiliation{\caltech}
\author{L.~J.~Broussard}\affiliation{\ornl}
\author{N.~B.~Callahan}\affiliation{\argonne}
\author{J.~H.~Choi}\affiliation{\ncsu}\affiliation{\tunl}
\author{S.~M.~Clayton}\affiliation{\lanl}
\author{S.~A.~Currie}\affiliation{\lanl}
\author{M.~Dawid}\affiliation{\indiana}\affiliation{\ceem}
\author{E.~B.~Dees}\affiliation{\ncsu}\affiliation{\tunl}
\author{B.~W.~Filippone}\affiliation{\caltech}
\author{W.~Fox}\affiliation{\indiana}\affiliation{\ceem}
\author{P.~Geltenbort}\affiliation{\ill}
\author{E.~George}\affiliation{\ttu}
\author{L.~Hayen}\affiliation{\ncsu}\affiliation{\tunl}
\author{K.~P.~Hickerson}\affiliation{\caltech}
\author{M.~A.~Hoffbauer}\affiliation{\lanl}
\author{K.~Hoffman}\affiliation{\ttu}
\author{A.~T.~Holley}\affiliation{\ttu}
\author{T.~M.~Ito}\affiliation{\lanl}
\author{A.~Komives}\affiliation{\depauw}
\author{C.-Y.~Liu}\affiliation{\indiana}\affiliation{\ceem}
\author{M.~Makela}\affiliation{\lanl}
\author{C.~L.~Morris}\affiliation{\lanl}
\author{R.~Musedinovic}\affiliation{\ncsu}\affiliation{\tunl}
\author{C.~O'Shaughnessy}\affiliation{\lanl}
\author{R.~W.~Pattie, Jr.}\affiliation{\etsu}
\author{J.~Ramsey}\affiliation{\ornl}
\author{D.~J.~Salvat}\affiliation{\indiana}\affiliation{\ceem}
\author{A.~Saunders}\affiliation{\lanl}\affiliation{\ornl}
\author{E.~I.~Sharapov}\affiliation{\jinr}
\author{S.~Slutsky}\affiliation{\caltech}
\author{V.~Su}\affiliation{\caltech}
\author{X.~Sun}\affiliation{\caltech}
\author{C.~Swank}\affiliation{\caltech}
\author{Z.~Tang}\affiliation{\lanl}
\author{W.~Uhrich}\affiliation{\lanl}
\author{J.~Vanderwerp}\affiliation{\indiana}\affiliation{\ceem}
\author{P.~Walstrom}\affiliation{\lanl}
\author{Z.~Wang}\affiliation{\lanl}
\author{W.~Wei}\affiliation{\caltech}
\author{A.~R.~Young}\affiliation{\ncsu}\affiliation{\tunl}
\collaboration{UCN$\tau$ collaboration}  
\date{\today}

\begin{abstract}
We report an improved measurement of the free neutron lifetime $\tau_{n}$ using the UCN$\tau$ apparatus at the Los Alamos Neutron Science Center. We count a total of approximately $38\times10^{6}$ surviving ultracold neutrons (UCN) after storing in UCN$\tau$'s magneto-gravitational trap over two data acquisition campaigns in 2017 and 2018. We extract $\tau_{n}$ from three blinded, independent analyses by both pairing long and short storage-time runs to find a set of replicate $\tau_{n}$ measurements and by performing a global likelihood fit to all data while self-consistently incorporating the $\beta$-decay lifetime. Both techniques achieve consistent results and find a value $\tau_{n}=877.75\pm0.28_{\text{ stat}}+0.22/-0.16_{\text{ syst}}$~s. With this sensitivity, neutron lifetime experiments now directly address the impact of recent refinements in our understanding of the standard model for neutron decay.
\end{abstract}

\maketitle
 
\paragraph{\label{sec:intro}Introduction.---} 

The decay $n\rightarrow p+e^{-}+\bar{\nu}_{e}$ is the simplest example of nuclear $\beta$-decay. The mean neutron lifetime $\tau_{n}$ provides a key input for predicting primordial light element abundances~\cite{J.Mathews2004}. The lifetime, combined with decay correlations, tests the $V$-$A$ structure of the weak interaction in the standard model (SM) at low energy while avoiding nuclear structure corrections~\cite{Dubbers2011}. Recent and forthcoming neutron $\beta$-decay experiments can be used to extract the magnitude of the Cabibbo-Kobayashi-Maskawa (CKM) matrix element $V_{ud}$ with a precision approaching that from studies of superallowed $0^{+}\rightarrow 0^{+}$ nuclear transitions. Furthermore, these tests can probe the existence of beyond-SM interactions at energy scales higher than 10~TeV which can evade detection at colliders~\cite{Ivanov2013,Gonzalez2019}.

The so-called inner electroweak radiative correction to $\tau_{n}$ contains model-dependent hadronic structure and short-distance QCD physics and is the dominant source of theoretical uncertainty in the prediction~\cite{Marciano2006}. A recent re-assessment of the inner correction using a data-driven dispersion relation approach shifts the value of $|V_{ud}|$ extracted from superallowed $0^{+}\rightarrow0^{+}$ decays resulting in a $\sim3\sigma$ deviation from unitarity in the top row of the CKM matrix~\cite{Seng2019,Czarnecki2019,Hayen2021}. This tension between the standard model and experiments has sparked a renewed urgency to independently examine these corrections. Studies of neutron decay, with increasing precision, are becoming a theoretically and experimentally robust standard model test~\cite{Czarnecki2019,Hayen2021}.

The current puzzle~\cite{Greene2016} due to the discrepancy between ``beam''~\cite{Yue2013} and ``bottle''~\cite{Serebrov2005,Pichlmaier2010,Steyerl2012,Arzumanov2015,Ezhov2018,Serebrov2018} methods of $\tau_{n}$ measurement indicates either the existence of new physics leading to a neutron decay channel without protons in the final state or the presence of inadequately assessed or unidentified systematic effects in either of the techniques. The former could be induced by the decay of neutrons to dark-matter particles~\cite{Fornal2018}, though such theories are constrained by astrophysical and cosmological signatures~\cite{Baym2018,McKeen2018,Ellis2018,McKeen2021A} and by direct searches for specific decay signatures~\cite{Sun2018,Tang2018}. Meanwhile, novel space-based and neutron-beam-based techniques aim to provide complementary $\tau_{n}$ measurements~\cite{Wilson2020,Lawrence2021,jparclifetime2019,Hirota2020}. This motivates a blinded measurement of $\tau_{n}$ with high statistical precision with multiple independent assessments of systematic effects and uncertainties.

In this Letter, we report a measurement of $\tau_{n}$ with $0.34$~s ($0.039$\%) uncertainty, improving upon our past results by a factor of $2.25$~\cite{Salvat2014,Morris2017,Pattie2018} using two blinded datasets from 2017 and 2018. The new result incorporates improved experimental and analysis techniques over Ref.~\cite{Pattie2018}. This is the first $\tau_{n}$ measurement precise enough to confront SM theoretical uncertainties. 

\paragraph{\label{sec:experiment}Experiment.---} 
    
\begin{figure}
    \centering
    \includegraphics[width=0.98\linewidth]{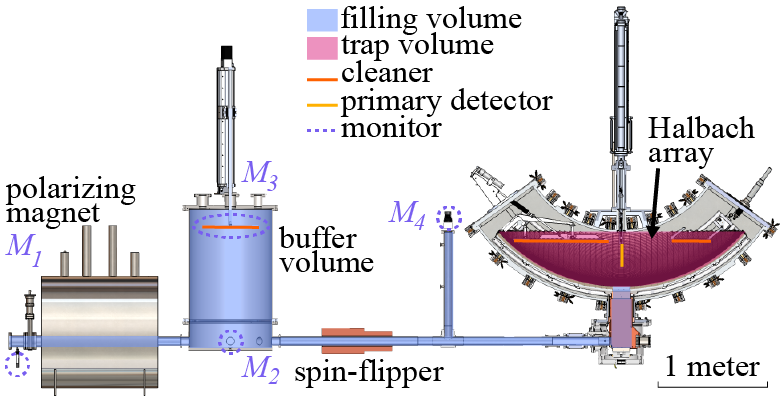}
    \caption{The UCN$\tau$ apparatus configuration for the 2018 campaign, with the volumes occupied by UCNs, cleaner surfaces, primary detector, and monitors highlighted. The polarizing magnet selects the high-field-seeking UCNs. UCNs are then ``precleaned'' in the buffer volume, spin-flipped to the trappable low-field-seeking spin state, and loaded into the trap. The 2017 configuration was the same as that of Ref. \cite{Pattie2018}.}
    \label{fig:apparatus}
\end{figure}

The experimental apparatus is depicted in Fig. \ref{fig:apparatus}. During a given run cycle, ultracold neutrons (UCNs) with kinetic energy $E\lesssim180$~neV from the Los Alamos Neutron Science Center's proton-beam-driven solid deuterium UCN source \cite{Ito2018} are transported to the UCN$\tau$ apparatus through a combination of NiP- and NiMo-coated guides. The UCNs are polarized by a $5.5$~T superconducting solenoid, spin-flipped to the trappable, low-field-seeking spin state via adiabatic fast-passage spin-flipper \cite{Holley2012}, and introduced into UCN$\tau$'s NdFeB bowl-shaped Halbach array \cite{Walstrom2009} over a time $t_{\text{load}}=150$~s ($300$~s) in the 2017 (2018) campaign. A $\sim15\times15$~cm$^{2}$ section of the Halbach array is then raised to close the bottom of the trap, magnetogravitationally confining the UCNs. A toroidal arrangement of electromagnetic coils provides a $60$-$120$~G ambient field to prevent UCN depolarization. UCNs with $E\gtrsim38$~neV are then removed (``cleaned'') \cite{Picker2009,Salvat2014,Morris2017,Pattie2018} during a period $t_{\text{clean}}=50$~s; they are either up-scattered by horizontal polyethylene sheets, or captured by $^{10}$B-coated-ZnS surfaces via the capture reaction $^{10}$B$+n\rightarrow\alpha+^{7}$Li. The cleaners are then retracted, and UCNs are stored in the trap for $t_{\text{store}}$ varying from $20$ to $1550$~s. The surviving neutrons are then counted by the primary $^{10}$B-coated-ZnS scintillator UCN counter \cite{Morris2017} over time interval $t_{\text{count}}=210$~s. The primary detector collects the $n$-capture-induced ZnS scintillation light in an array of wavelength shifting fibers, which route the light to two photomultiplier tubes (PMTs). The counting phase occurs in three stages, with the detector first lowered just into the trap, but at the height of the cleaners for $40$~s, then lowered into the middle for $20$~s, and finally lowered to the bottom for $150$~s. This procedure constrains both the number of remaining uncleaned UCNs and the presence of $t_{\text{store}}$-dependent changes to the energy spectrum of the trapped UCNs.

The smaller cleaner (rightmost in Fig. \ref{fig:apparatus}) was viewed by four PMTs, and scintillation light from the UCN captured by the $^{10}$B coating were counted during the cleaning process similar to the primary counter. Further, this cleaner was relowered during the counting phase to search for any uncleaned higher-energy UCNs.

A buffer volume was introduced in 2018 that precleans the loaded UCNs and smooths over any temporal fluctuations in the UCN production rate while loading. 

Run cycles are performed in short ($t_{\text{store}}\leq 500$~s) and long ($t_{\text{store}}>500$~s) pairs. Backgrounds are measured by performing runs with the same sequencing but with no protons on target, with protons on target but the UCN source valved off from UCN$\tau$, and with the UCN$\tau$ trap door closed. Additional background runs are acquired during facility downtime. These latter background runs measure the vertical position-dependent primary counter background variations, likely due to the position-dependent probability of ZnS scintillation light reflecting from the apparatus into the PMTs.

UCNs within the loading volume are counted by a set of monitor detectors $M_{1}$-$M_{4}$ (Fig. \ref{fig:apparatus}). The monitor detectors are $^{10}$B-coated-ZnS sheets coupled directly to PMTs~\cite{Wang2015}. The monitor detectors provide data to normalize the primary detector counts, correcting for variations in UCN source intensity and the energy spectrum of UCNs. Detectors at different heights relative to the beam line have differing sensitivity to the energy spectrum, and analyzing the full ensemble of monitor detectors captures changes to the spectrum. For example, heat load on the UCN source during operation gradually reduces the solid deuterium crystal quality, hardening the spectrum~\cite{Anghel2018}. This changes the relationship between the monitor counts and the number of initially-trapped UCNs in a given run. We periodically melt and refreeze the D$_{2}$ source to restore source quality

The single photoelectron (PE) primary counter PMT signals are split into low ($\sim1/6$ PE) and high ($\sim1/3$ PE) threshold channels and discriminated with $16$~ns dead time. Monitor detector signals exhibit higher light yield per neutron, and are shaped and discriminated with a threshold that selects UCN captures and rejects background~\cite{Wang2015}. All channels are read by a $1.25$~GHz multichannel scaler. Data are blinded as in Ref. \cite{Pattie2018}, shifting $\tau_{n}$ by an unknown factor within a $\pm15$~s window.
    
\paragraph{\label{sec:analysis}Analysis.---} 
We performed three analyses $A$, $B$, and $C$. Initially, no extracted values of $\tau_{n}$ were shared between analyzers so that analyses developed separately. After sharing blinded values, we performed run-by-run comparisons of extracted monitor counts, primary detector counts, and background estimates between analyzers. The data were unblinded when the three extracted $\tau_{n}$ values agreed to within $0.1$~s. Each analysis separately assessed systematic effects and potential statistical biases.

\begin{figure}
    \centering
    \includegraphics[width=1.0\linewidth,trim={0cm 0cm 1cm 1.0cm},clip]{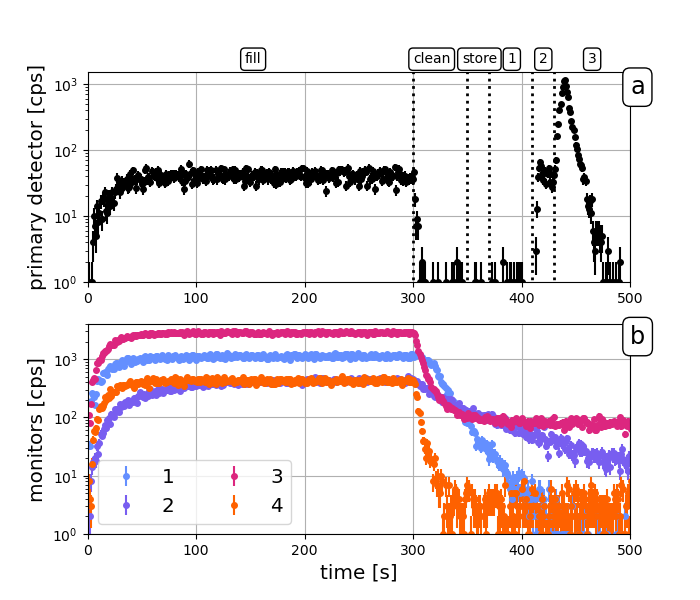}
    \caption{(a) Primary detector coincidence counts from analysis $A$ for a typical $t_{\mathrm{store}}=20$~s run cycle. From left to right, the dotted vertical lines mark the time windows for filling, cleaning, storing, and three counting stages. (b) The count rate during the run cycle for the monitors deployed in the 2018 analysis.}
    \label{fig:data_example}
\end{figure}

All analyses proceeded in the following stages. Run quality criteria are developed to reject runs with poor or abnormal experimental conditions. For each remaining run $j$, primary detector events are formed. These events and monitor counts are time binned as in Fig. \ref{fig:data_example}, and rate-dependent corrections are applied to counts in each bin. Corrected primary detector counts $d_{j}(t)$ are summed during the counting phase to give an UCN signal $D_{j}$. A combination of the rates from background runs proximal in time to $j$, as well as counts during storage and after counting, are used to estimate background counts $B_{j}$ during the three counting stages. The monitor detector counts $m_{k,j}(t)$ are time weighted by a function $w(t)$ and summed for $0<t<t_{\text{fill}}$ to find monitor signals $M_{k,j}$. These are reduced to a single normalization factor using a regression model $f\left(M_{k,j}\right)\rightarrow N_{j}$. The quantities $D_{j}$, $B_{j}$, and $N_{j}$ are then used to find $\tau_{n}$ as discussed below.

Analyses identified and removed runs with significant fluctuations of the LANSCE proton beam during filling and excessive electronic noise while moving the cleaners and primary detector.

An UCN event in the primary detector is characterized by the presence of coincident single PEs between the two PMTs within time $t_{c}$ (ranging from $50$ to $100$~ns among analyzers). Thereafter, subsequent PE are counted for a time $t_{w}$ ranging from $600$ to $1400$~ns, and this time window extended if additional PEs are found in the current window. Total counting times vary between $t_{\text{event}}\sim500$ and $4000$~ns. Analysis $A$ ($B$) requires $n_{p}\geq8$ $(6)$ PEs in an initial $1000$~ns ($600$~ns) window. All analyses varied these parameters to test the sensitivity of $\tau_{n}$. Reference \cite{Pattie2018} used the raw PE rates (referred to as a ``singles analysis'' therein) to determine the number of UCNs, which had the advantage of smaller rate-dependent effects and the disadvantage of poorer signal to background. This treatment is inappropriate for the current data due to slow drifts in single PE background rates over the course of the measurement campaign.

This coincidence method gives signal to background for short $t_{\text{store}}$ runs as high as 10$^{3}$. However, ZnS[Ag]'s long scintillation time constants makes this method subject to rate-dependent systematic effects from dead time, pileup, and accidental event retriggering due to delayed scintillation light. Analysis $C$ maintained a time-averaged rate from previous UCN events within a given run and varied $n_{p}$ event by event using the known time distribution of the ZnS scintillation. This reduces the possibility of a previous event contributing delayed PEs to the current event, thereby selecting an otherwise underthreshold event. Upon time binning events during $t_{\text{count}}$, analysis $C$ tabulates the sum of $t_{\text{event}}$ in each bin and performs an analytic correction to the number of counts in the bin as a function of instantaneous event rate. Analyses $A$ and $B$ used the measured ZnS scintillation time dependence and typical UCN count rates to generate pseudo-datasets. These pseudo-data were analyzed and compared to the true number of simulated counts to develop corrections for rate-dependent effects.

\begin{table*}[t]
    \centering
    \begin{tabular}{c||c|c|c}
        Analysis & $w(t)$ & $f$ & Comments \\ \hline\hline
        $A$ & $\exp\left(\kappa t\right)$ & $c_{1\text{A}} M_{l,j} + c_{2\text{A}} M_{u,j}$ & $l=1$(2) and $u=4$(3) in 2017 (2018) \\ \hline
        $B$ & $\left[1+\exp{\left(-\frac{t-\left(t_{\text{fill}}/\alpha\right)}{t_{\text{fill}}/\beta}\right)}\right]^{-1}$ & $c_{0B}+c_{1\text{B}}\tilde{M}_{1,j}+c_{2\text{B}}\tilde{M}_{2,j}$ & $\alpha=2$, $\beta=10$, $\tilde{M}_{k,j}$ from PCA of $z$-scored $M_{k,j}$ \\ \hline
        $C$ & $\sum_{m=1}^{20}A_{m}U\left(t|(m-1)t_{\text{fill}},mt_{\text{fill}}\right)$ & $c_{1\text{C}} M_{l,j} + c_{2\text{C}} M_{u,j}$ & ${A_{m}}$ chosen to minimize variance in $M_{k,j}$ \\ \hline
    \end{tabular}
    \caption{The monitor detector weighting functions $w(t)$ and function $f(M_{k,j})$ used to find run-by-run normalization factor $N_{j}$. Analysis $B$ $z$-scores the $M_{k,j}$ over a range of $j$ and performs a principal component analysis (PCA) to find the two dominant principal components $\tilde{M}_{1/2,j}$. $U(t|a,b)$ is the uniform distribution over $(a,b)$. Note that the coefficients $c$ vary over different subsets of runs (see text).}
    \label{tab:normalization}
\end{table*}

The choices of $w(t)$ and $f(M_{k,j})$ for each analysis are given in Table \ref{tab:normalization}. In general, the given choices of $w(t)$ deweight counts early in the filling phase, which better predicts the number of trapped UCNs. Indeed, because the duration of the filling phase is several times longer than the characteristic time for UCNs to fill the guides and apparatus, a fluctuation in UCN source output occurring early in the filling phase is less likely to influence the number occupying the trap by time $t_{\text{fill}}$. Runs are generally grouped into week- or month-long subsets between melt and refreezes and other discrete changes to the apparatus or the source. The coefficients $c$ thus vary among subsets. Analysis $B$ uses a principal component analysis of the $z$-scored monitor signals for all monitors~\cite{principlecomponents}. In practice, two components dominate and are used in the normalization: the first component tracks the coarse, overall changes in source output, while the second captures changes in the relative amount of trappable lower-energy UCNs compared to untrappable higher-energy UCNs.

As was done in Ref. \cite{Pattie2018}, all analyses use the mean arrival time $\bar{t}_{j}=\int t d_{j}(t) dt$ over the counting phase rather than the nominal $t_{\text{store}}$. This corrects for potential run-by-run or $t_{\text{store}}$-dependent variation of the characteristic timescale with which UCNs are absorbed by the primary detector during counting. Using the nominal $t_{\text{store}}$ shifts $\tau_{n}$ by $0.01$--$0.03$~s.

\begin{table}[ht]
    \centering
    \begin{tabular}{c|rr}
        Effect & Correction & uncertainty \\
        \hline\hline
        UCN event definition      & ...     & $\pm0.13$ \\
        Normalization weighting   & ...     & $\pm0.06$ \\
        Depolarization            & ...     & $+0.07$   \\
        Uncleaned UCN             & ...     & $+0.11$   \\
        Heated UCN                & ...     & $+0.08$   \\
        Al block                  & $+0.06$ & $\pm0.05$ \\
        Residual gas scattering   & $+0.11$ & $\pm0.06$ \\
        \hline
        Uncorrelated sum & \multicolumn{2}{c}{$0.17^{+0.22}_{-0.16}$ s} \\
        \hline
    \end{tabular}
    \caption{Systematic corrections with uncertainties to $\tau_{n}$ in seconds. The total is an uncorrelated combination of all systematic corrections. The non-zero corrections here are applied to the final result. Only the UCN event definition and normalization effects can decrease the measured $\tau_{n}$.}
    \label{tab:systematics}
\end{table}

\paragraph{\label{sec:tau_n}Extraction of $\tau_{n}$.---} 
We determine $\tau_{n}$ by both a ``paired'' and a ``global'' analysis. In the former, pairs of short ($j\equiv s$) and long ($j\equiv l$) storage time runs proximal in time are used to compute
\begin{equation}
    \tau_{n}^{(p)} = \frac{\bar{t}_{l}-\bar{t}_{s}}{\log\left(Y_{s}/Y_{l}\right)}
\end{equation}
with $Y_{j}=\left(D_{j}-B_{j}\right)/N_{j}$, and uncertainties propagated to find $\delta\tau_{n}^{(p)}$. A bias-corrected weighted average over pairs $p$ gives the final result. This method is less sensitive to potential time drifts in backgrounds or detector response. However, the paired method is less statistically sensitive to $\tau_{n}$ than the global method discussed below, as the pairing process reduces the number of runs used in the analyses. This method finds $\tau_{n}$ in agreement with that of the global analysis.

The global method instead maximizes the likelihood of observing the data $\mathbf{O}=\left\{D_{j},B_{j},M_{k,j},\bar{t}_{j}\right\}$ given parameters $\tau_{n}$, $\mathbf{p}$:
\begin{eqnarray}
    \mathcal{L}\left(\mathbf{O}|\tau_{n},\mathbf{p}\right) &=& \prod_{j}\mathcal{P}\left(\lambda_{j},\mathbf{p}|\textbf{O}\right)\prod_{h} \pi_{h,j}\left(p_{h,j}\right)
\end{eqnarray}
with $\lambda_{j}=\tau_{n}^{-1}+\lambda_{\text{up},j}+\lambda_{\text{dp}}+\lambda_{\text{escape}}$ the loss rate including $\beta$ decay, residual gas up-scattering rate $\lambda_{\text{up},j}$, depolarization rate $\lambda_{\text{dp}}$, and escape rate $\lambda_{\text{escape}}$ described below. The probability distribution $\mathcal{P}$, the collection of nuisance parameters $\mathbf{p}$, and their constraints $\pi(p_{h})$ vary between analyses. Analysis $A$ computed the Poisson-distributed probability of observing the primary detector counts $D_{j}$ for a predicted mean number of counts $N_{j}\exp(-\lambda\bar{t}_{j})+B_{j}$. The \texttt{EMCEE} Markov-chain Monte Carlo algorithm~\cite{ForemanMackey2013} was used to marginalize over $c_{1A,j}$, $c_{2A,j}$, and $B_{j}$ to find $\tau_{n}$ with uncertainty~\cite{GonzalezDissertation}. Analysis $B$ modeled the $D_{j}$ as quasi-Poisson distributed~\cite{VerHoef2007} with the parameters in the normalization model and extra scaling parameter in the variance to account for non-statistical fluctuations treated as free nuisance parameters. A profile likelihood provides $\tau_{n}$ with uncertainty~\cite{FriesDissertation}. Analysis $C$ instead minimizes the $\chi^{2}$ of the yields $Y_{j}$ with uncertainties over $\tau_{n}$, normalization parameters, and a free scaling parameter in the variance of the yields. Correlations in the parameter uncertainties are incorporated to find $\tau_{n}$ with uncertainty. 

\paragraph{Systematic effects.---} All systematic corrections and their uncertainties are summarized in Table \ref{tab:systematics}. During 2017, accounting for $\sim34$\% of the full dataset, a block-shaped $2.5\times2.5\times1.3$~cm$^{3}$ aluminum structural component from one of the cleaning surfaces fell into the trapping volume, leading to an additional loss channel for UCNs due to up-scattering and absorption on the Al surface. The effect of the aluminum block's additional material loss was assessed in a separate, dedicated run wherein the block was covered in polyethylene foil and reinserted into the same location in the trap. Polyethylene up scatters virtually all UCNs incident upon the surface, thus providing an estimate of the rate of UCNs impinging upon the aluminum block. This introduces an additional $\sim424$~s decay component to the yields $Y_{j}$ versus $\bar{t}_{j}$. Analysis $B$ uses an estimate of UCN velocity-and-angle-averaged loss per bounce of $(2\pm1)\times10^{-4}$ for aluminum, leading to a $0.15\pm0.07$~s correction to this subset of data. The other analyses perform similar assessments, and independent checks using our Monte Carlo framework \cite{Callahan2019} are consistent with this estimate. Further, this is consistent with a comparison of paired $\tau_{n}$ during this period to the rest of the paired $\tau_{n}$ values. The value in Table \ref{tab:systematics} is this correction's contribution to the final $\tau_{n}$ for all data.

If UCNs with $E>51$~neV (corresponding to the $50$~cm height of the trap) are not removed during the cleaning phase, and yet reside in the trap for times $\lambda_{\text{escape}}^{-1}$ similar to $\tau_{n}$ before escaping, the extracted $\tau_{n}$ will be systematically lower. Data were obtained in 2018 wherein the cleaner was never lowered into the trap. For these runs, the cleaner detector and primary detector both register counts at positions above the cleaning height during the counting phase, and the extracted $\tau_{n}$ lowered by $\sim15$~s, implying an escape rate $\lambda_{\text{escape}}=2\times10^{-5}$~s$^{-1}$. Assuming a linear relation between $\lambda_{\text{escape}}$ and observed counts in either the cleaner or primary detector, we use the nonobservation of counts in the first counting phase in the primary datasets to constrain the size of the effect. Similarly, as was done in Ref. \cite{Pattie2018}, the nonobservation of counts at the highest primary detector position for $t_{\text{store}}>500$~s runs in the primary data are used to constrain the effect of initially trappable UCNs, increasing their total energy due to vibration of the Halbach array or any other time dependence in the trapping potential. 

The choice of UCN event definition can change the extracted $\tau_{n}$. Each analysis reextracted $\tau_{n}$ for a range of $n_{p}$, $t_{w}$, and any other adjustable parameters to assess the associated systematic uncertainty. The three analyses observe a variance similar in magnitude (see Table \ref{tab:systematics}).

The sensitivity of the extracted lifetime to choice of weighting function $w(t)$ was assessed by defining appropriate model parameter ranges consistent with accurately modeling the initially loaded UCN populations (e.g., $\alpha$, $\beta$ for analysis $B$), and varying the parameters within these ranges to probe the effect on the extracted lifetime. Taking the most conservative assessment among analyses results in an uncertainty in the lifetime of $0.090$~s in 2017 and $0.047$~s in 2018 for a combined uncertainty $0.06$~s. The nearly $50$\% reduction in uncertainty in 2018 is largely due to the buffer volume smoothing over variations in UCN production while filling.

Depolarization and residual gas up-scattering losses and associated uncertainties were treated as in Ref. \cite{Pattie2018}. The toroidal ambient magnetic field strength is unchanged from Ref. \cite{Pattie2018}, and we assign a depolarization rate of $\lambda_{\text{dp}}=0.0^{+1.0}_{-0.0}\times10^{-7}$~s$^{-1}$. The residual gas up-scattering rate was computed run by run using the measured absolute pressure and periodic residual gas analysis of the trap. The relative amounts of water, air, and hydrocarbon molecules were tracked throughout the run campaigns. The up-scattering rate is computed from these data and from the known UCN-gas molecule cross sections measured in Refs.~\cite{Seestrom2015,Seestrom2017}. We assign a conservative $15$\% uncertainty on the absolute pressure and cross section uncertainty on water and air from the uncertainties in Refs. \cite{Seestrom2015,Seestrom2017}. Because of uncertainty in the relative abundance of hydrocarbons, we apply a $70$\% uncertainty, which spans the full range of measured cross sections in Refs.~\cite{Seestrom2015,Seestrom2017}.

\paragraph{\label{sec:results}Results and conclusions.---}

\begin{table}[ht]
    \centering
    \begin{tabular}{l|c|c|c}
        Analysis & $A$ & $B$ & $C$ \\
        \hline\hline
        Paired & $877.77\pm0.31$ & $877.87\pm0.32$ & $877.60\pm0.29$ \\
        \hline
        2017 & $877.68\pm0.30$ & $877.78\pm0.34$ & $877.74\pm0.33$\\
        2018 & $878.06\pm0.49$ & $877.80\pm0.46$ & $877.55\pm0.55$ \\
        With Al & $877.17\pm0.38$ & $876.81\pm0.43$ & $877.69\pm0.40$ \\
        Full & $877.78\pm0.26$ & $877.79\pm0.27$ & $877.69\pm0.28$ \\
        \hline
        Combined &  \multicolumn{3}{c}{$877.75\pm0.28$} \\
        \hline
    \end{tabular}
    \caption{The resulting $\tau_{n}$ (s) for the three analyses with statistical uncertainties. The first row shows paired results for the full dataset. Below are the global fit results for the 2017 and 2018 subsets, and the subset for which the Al component was present in the trap. ``Full'' refers to the global maximum likelihood fit of all production data. An unweighted average of the three global analyses and largest uncertainty gives the final result.}
    \label{tab:combined_result}
\end{table}

The extracted $\tau_{n}$ values with statistical uncertainties for 2017 and 2018 for the different analyses are shown in Table \ref{tab:combined_result}. Table \ref{tab:systematics} shows the systematic corrections and most conservative uncertainties among the three analyses. Because of differing run selection, UCN event definition, normalization models, and background estimates, the statistical uncertainties between the analyses are not $100$\% correlated. We explored multiple ways of averaging the three analyses, with little benefit or difference when using optimized estimator averaging procedures (similar to Ref. \cite{Albahri2021}). Ultimately, as in our previous result \cite{Pattie2018}, we perform an unweighted average of the three central values and choose the largest statistical uncertainty, giving $\tau_{n}=877.75\pm0.28_{\text{ stat}}+0.22/-0.16_{\text{ syst}}$~s, in agreement with Ref. \cite{Pattie2018}.

With $0.039$\% total uncertainty, this is the first experiment to determine the neutron lifetime with uncertainty smaller than the average $0.09(2)$\% shift predicted from recent theoretical work \cite{Seng2019,Czarnecki2019,Hayen2021} on radiative corrections. This result, combined with forthcoming experimental and lattice QCD determinations of the axial coupling constant, will independently probe the disagreement in first-row CKM unitarity while avoiding nuclear structure corrections required for superallowed nuclear decays. 

We thank W. Pettus for a thorough reading of this manuscript. This work is supported by the LANL LDRD program; the U.S. Department of Energy, Office of Science, Office of Nuclear Physics under Awards No. DE-FG02-ER41042, No. DE-AC52-06NA25396, No. DE-AC05-00OR2272, and No. 89233218CNA000001 under proposal LANLEEDM; NSF Grants No. 1614545, No. 1914133, No. 1506459, No. 1553861, No. 1812340, No. 1714461, and No. 1913789; and NIST precision measurements grant.

\bibliography{main.bib}

\begin{thebibliography}{42}%
\makeatletter
\providecommand \@ifxundefined [1]{%
 \@ifx{#1\undefined}
}%
\providecommand \@ifnum [1]{%
 \ifnum #1\expandafter \@firstoftwo
 \else \expandafter \@secondoftwo
 \fi
}%
\providecommand \@ifx [1]{%
 \ifx #1\expandafter \@firstoftwo
 \else \expandafter \@secondoftwo
 \fi
}%
\providecommand \natexlab [1]{#1}%
\providecommand \enquote  [1]{``#1''}%
\providecommand \bibnamefont  [1]{#1}%
\providecommand \bibfnamefont [1]{#1}%
\providecommand \citenamefont [1]{#1}%
\providecommand \href@noop [0]{\@secondoftwo}%
\providecommand \href [0]{\begingroup \@sanitize@url \@href}%
\providecommand \@href[1]{\@@startlink{#1}\@@href}%
\providecommand \@@href[1]{\endgroup#1\@@endlink}%
\providecommand \@sanitize@url [0]{\catcode `\\12\catcode `\$12\catcode
  `\&12\catcode `\#12\catcode `\^12\catcode `\_12\catcode `\%12\relax}%
\providecommand \@@startlink[1]{}%
\providecommand \@@endlink[0]{}%
\providecommand \url  [0]{\begingroup\@sanitize@url \@url }%
\providecommand \@url [1]{\endgroup\@href {#1}{\urlprefix }}%
\providecommand \urlprefix  [0]{URL }%
\providecommand \Eprint [0]{\href }%
\providecommand \doibase [0]{http://dx.doi.org/}%
\providecommand \selectlanguage [0]{\@gobble}%
\providecommand \bibinfo  [0]{\@secondoftwo}%
\providecommand \bibfield  [0]{\@secondoftwo}%
\providecommand \translation [1]{[#1]}%
\providecommand \BibitemOpen [0]{}%
\providecommand \bibitemStop [0]{}%
\providecommand \bibitemNoStop [0]{.\EOS\space}%
\providecommand \EOS [0]{\spacefactor3000\relax}%
\providecommand \BibitemShut  [1]{\csname bibitem#1\endcsname}%
\let\auto@bib@innerbib\@empty
\bibitem [{\citenamefont {J.~Mathews}\ \emph {et~al.}(2004)\citenamefont
  {J.~Mathews}, \citenamefont {Kajino},\ and\ \citenamefont
  {Shima}}]{J.Mathews2004}%
  \BibitemOpen
  \bibfield  {author} {\bibinfo {author} {\bibfnamefont {G.}~\bibnamefont
  {J.~Mathews}}, \bibinfo {author} {\bibfnamefont {T.}~\bibnamefont {Kajino}},
  \ and\ \bibinfo {author} {\bibfnamefont {T.}~\bibnamefont {Shima}},\ }\href
  {\doibase 10.1103/PhysRevD.71.021302} {\bibfield  {journal} {\bibinfo
  {journal} {arXiv [astro-ph]}\ } (\bibinfo {year} {2004}),\
  10.1103/PhysRevD.71.021302}\BibitemShut {NoStop}%
\bibitem [{\citenamefont {Dubbers}\ and\ \citenamefont
  {Schmidt}(2011)}]{Dubbers2011}%
  \BibitemOpen
  \bibfield  {author} {\bibinfo {author} {\bibfnamefont {D.}~\bibnamefont
  {Dubbers}}\ and\ \bibinfo {author} {\bibfnamefont {M.~G.}\ \bibnamefont
  {Schmidt}},\ }\href {\doibase 10.1103/RevModPhys.83.1111} {\bibfield
  {journal} {\bibinfo  {journal} {Rev. Mod. Phys.}\ }\textbf {\bibinfo {volume}
  {83}},\ \bibinfo {pages} {1111} (\bibinfo {year} {2011})}\BibitemShut
  {NoStop}%
\bibitem [{\citenamefont {Ivanov}\ \emph {et~al.}(2013)\citenamefont {Ivanov},
  \citenamefont {Pitschmann},\ and\ \citenamefont {Troitskaya}}]{Ivanov2013}%
  \BibitemOpen
  \bibfield  {author} {\bibinfo {author} {\bibfnamefont {A.~N.}\ \bibnamefont
  {Ivanov}}, \bibinfo {author} {\bibfnamefont {M.}~\bibnamefont {Pitschmann}},
  \ and\ \bibinfo {author} {\bibfnamefont {N.~I.}\ \bibnamefont {Troitskaya}},\
  }\href {\doibase 10.1103/PhysRevD.88.073002} {\bibfield  {journal} {\bibinfo
  {journal} {Phys. Rev. D}\ }\textbf {\bibinfo {volume} {88}},\ \bibinfo
  {pages} {073002} (\bibinfo {year} {2013})}\BibitemShut {NoStop}%
\bibitem [{\citenamefont {Gonzalez-Alonso}\ \emph {et~al.}(2019)\citenamefont
  {Gonzalez-Alonso}, \citenamefont {Naviliat-Cuncic},\ and\ \citenamefont
  {Severijns}}]{Gonzalez2019}%
  \BibitemOpen
  \bibfield  {author} {\bibinfo {author} {\bibfnamefont {M.}~\bibnamefont
  {Gonzalez-Alonso}}, \bibinfo {author} {\bibfnamefont {O.}~\bibnamefont
  {Naviliat-Cuncic}}, \ and\ \bibinfo {author} {\bibfnamefont {N.}~\bibnamefont
  {Severijns}},\ }\href {\doibase https://doi.org/10.1016/j.ppnp.2018.08.002}
  {\bibfield  {journal} {\bibinfo  {journal} {Progress in Particle and Nuclear
  Physics}\ }\textbf {\bibinfo {volume} {104}},\ \bibinfo {pages} {165 }
  (\bibinfo {year} {2019})}\BibitemShut {NoStop}%
\bibitem [{\citenamefont {Marciano}\ and\ \citenamefont
  {Sirlin}(2006)}]{Marciano2006}%
  \BibitemOpen
  \bibfield  {author} {\bibinfo {author} {\bibfnamefont {W.~J.}\ \bibnamefont
  {Marciano}}\ and\ \bibinfo {author} {\bibfnamefont {A.}~\bibnamefont
  {Sirlin}},\ }\href {\doibase 10.1103/PhysRevLett.96.032002} {\bibfield
  {journal} {\bibinfo  {journal} {Phys. Rev. Lett.}\ }\textbf {\bibinfo
  {volume} {96}},\ \bibinfo {pages} {032002} (\bibinfo {year}
  {2006})}\BibitemShut {NoStop}%
\bibitem [{\citenamefont {Seng}\ \emph {et~al.}(2019)\citenamefont {Seng},
  \citenamefont {Gorchtein},\ and\ \citenamefont {Ramsey-Musolf}}]{Seng2019}%
  \BibitemOpen
  \bibfield  {author} {\bibinfo {author} {\bibfnamefont {C.-Y.}\ \bibnamefont
  {Seng}}, \bibinfo {author} {\bibfnamefont {M.}~\bibnamefont {Gorchtein}}, \
  and\ \bibinfo {author} {\bibfnamefont {M.~J.}\ \bibnamefont
  {Ramsey-Musolf}},\ }\href {\doibase 10.1103/PhysRevD.100.013001} {\bibfield
  {journal} {\bibinfo  {journal} {Phys. Rev. D}\ }\textbf {\bibinfo {volume}
  {100}},\ \bibinfo {pages} {013001} (\bibinfo {year} {2019})}\BibitemShut
  {NoStop}%
\bibitem [{\citenamefont {Czarnecki}\ \emph {et~al.}(2019)\citenamefont
  {Czarnecki}, \citenamefont {Marciano},\ and\ \citenamefont
  {Sirlin}}]{Czarnecki2019}%
  \BibitemOpen
  \bibfield  {author} {\bibinfo {author} {\bibfnamefont {A.}~\bibnamefont
  {Czarnecki}}, \bibinfo {author} {\bibfnamefont {W.~J.}\ \bibnamefont
  {Marciano}}, \ and\ \bibinfo {author} {\bibfnamefont {A.}~\bibnamefont
  {Sirlin}},\ }\href {\doibase 10.1103/PhysRevD.100.073008} {\bibfield
  {journal} {\bibinfo  {journal} {Phys. Rev. D}\ }\textbf {\bibinfo {volume}
  {100}},\ \bibinfo {pages} {073008} (\bibinfo {year} {2019})}\BibitemShut
  {NoStop}%
\bibitem [{\citenamefont {Hayen}(2021)}]{Hayen2021}%
  \BibitemOpen
  \bibfield  {author} {\bibinfo {author} {\bibfnamefont {L.}~\bibnamefont
  {Hayen}},\ }\href {\doibase 10.1103/PhysRevD.103.113001} {\bibfield
  {journal} {\bibinfo  {journal} {Phys. Rev. D}\ }\textbf {\bibinfo {volume}
  {103}},\ \bibinfo {pages} {113001} (\bibinfo {year} {2021})}\BibitemShut
  {NoStop}%
\bibitem [{\citenamefont {Greene}\ and\ \citenamefont
  {Geltenbort}(2016)}]{Greene2016}%
  \BibitemOpen
  \bibfield  {author} {\bibinfo {author} {\bibfnamefont {G.~L.}\ \bibnamefont
  {Greene}}\ and\ \bibinfo {author} {\bibfnamefont {P.}~\bibnamefont
  {Geltenbort}},\ }\href@noop {} {\bibfield  {journal} {\bibinfo  {journal}
  {Scientific American}\ }\textbf {\bibinfo {volume} {314}} (\bibinfo {year}
  {2016})}\BibitemShut {NoStop}%
\bibitem [{\citenamefont {Yue}\ \emph {et~al.}(2013)\citenamefont {Yue},
  \citenamefont {Dewey}, \citenamefont {Gilliam}, \citenamefont {Greene},
  \citenamefont {Laptev}, \citenamefont {Nico}, \citenamefont {Snow},\ and\
  \citenamefont {Wietfeldt}}]{Yue2013}%
  \BibitemOpen
  \bibfield  {author} {\bibinfo {author} {\bibfnamefont {A.~T.}\ \bibnamefont
  {Yue}}, \bibinfo {author} {\bibfnamefont {M.~S.}\ \bibnamefont {Dewey}},
  \bibinfo {author} {\bibfnamefont {D.~M.}\ \bibnamefont {Gilliam}}, \bibinfo
  {author} {\bibfnamefont {G.~L.}\ \bibnamefont {Greene}}, \bibinfo {author}
  {\bibfnamefont {A.~B.}\ \bibnamefont {Laptev}}, \bibinfo {author}
  {\bibfnamefont {J.~S.}\ \bibnamefont {Nico}}, \bibinfo {author}
  {\bibfnamefont {W.~M.}\ \bibnamefont {Snow}}, \ and\ \bibinfo {author}
  {\bibfnamefont {F.~E.}\ \bibnamefont {Wietfeldt}},\ }\href {\doibase
  10.1103/PhysRevLett.111.222501} {\bibfield  {journal} {\bibinfo  {journal}
  {Phys. Rev. Lett.}\ }\textbf {\bibinfo {volume} {111}},\ \bibinfo {pages}
  {222501} (\bibinfo {year} {2013})}\BibitemShut {NoStop}%
\bibitem [{\citenamefont {Serebrov}\ \emph {et~al.}(2005)\citenamefont
  {Serebrov}, \citenamefont {Varlamov}, \citenamefont {Kharitonov},
  \citenamefont {Fomin}, \citenamefont {Pokotilovski}, \citenamefont
  {Geltenbort}, \citenamefont {Butterworth}, \citenamefont {Krasnoschekova},
  \citenamefont {Lasakov}, \citenamefont {Tal'daev}, \citenamefont
  {Vassiljev},\ and\ \citenamefont {Zherebtsov}}]{Serebrov2005}%
  \BibitemOpen
  \bibfield  {author} {\bibinfo {author} {\bibfnamefont {A.}~\bibnamefont
  {Serebrov}}, \bibinfo {author} {\bibfnamefont {V.}~\bibnamefont {Varlamov}},
  \bibinfo {author} {\bibfnamefont {A.}~\bibnamefont {Kharitonov}}, \bibinfo
  {author} {\bibfnamefont {A.}~\bibnamefont {Fomin}}, \bibinfo {author}
  {\bibfnamefont {Y.}~\bibnamefont {Pokotilovski}}, \bibinfo {author}
  {\bibfnamefont {P.}~\bibnamefont {Geltenbort}}, \bibinfo {author}
  {\bibfnamefont {J.}~\bibnamefont {Butterworth}}, \bibinfo {author}
  {\bibfnamefont {I.}~\bibnamefont {Krasnoschekova}}, \bibinfo {author}
  {\bibfnamefont {M.}~\bibnamefont {Lasakov}}, \bibinfo {author} {\bibfnamefont
  {R.}~\bibnamefont {Tal'daev}}, \bibinfo {author} {\bibfnamefont
  {A.}~\bibnamefont {Vassiljev}}, \ and\ \bibinfo {author} {\bibfnamefont
  {O.}~\bibnamefont {Zherebtsov}},\ }\href {\doibase
  https://doi.org/10.1016/j.physletb.2004.11.013} {\bibfield  {journal}
  {\bibinfo  {journal} {Physics Letters B}\ }\textbf {\bibinfo {volume}
  {605}},\ \bibinfo {pages} {72 } (\bibinfo {year} {2005})}\BibitemShut
  {NoStop}%
\bibitem [{\citenamefont {Pichlmaier}\ \emph {et~al.}(2010)\citenamefont
  {Pichlmaier}, \citenamefont {Varlamov}, \citenamefont {Schreckenbach},\ and\
  \citenamefont {Geltenbort}}]{Pichlmaier2010}%
  \BibitemOpen
  \bibfield  {author} {\bibinfo {author} {\bibfnamefont {A.}~\bibnamefont
  {Pichlmaier}}, \bibinfo {author} {\bibfnamefont {V.}~\bibnamefont
  {Varlamov}}, \bibinfo {author} {\bibfnamefont {K.}~\bibnamefont
  {Schreckenbach}}, \ and\ \bibinfo {author} {\bibfnamefont {P.}~\bibnamefont
  {Geltenbort}},\ }\href {\doibase
  https://doi.org/10.1016/j.physletb.2010.08.032} {\bibfield  {journal}
  {\bibinfo  {journal} {Physics Letters B}\ }\textbf {\bibinfo {volume}
  {693}},\ \bibinfo {pages} {221 } (\bibinfo {year} {2010})}\BibitemShut
  {NoStop}%
\bibitem [{\citenamefont {Steyerl}\ \emph {et~al.}(2012)\citenamefont
  {Steyerl}, \citenamefont {Pendlebury}, \citenamefont {Kaufman}, \citenamefont
  {Malik},\ and\ \citenamefont {Desai}}]{Steyerl2012}%
  \BibitemOpen
  \bibfield  {author} {\bibinfo {author} {\bibfnamefont {A.}~\bibnamefont
  {Steyerl}}, \bibinfo {author} {\bibfnamefont {J.~M.}\ \bibnamefont
  {Pendlebury}}, \bibinfo {author} {\bibfnamefont {C.}~\bibnamefont {Kaufman}},
  \bibinfo {author} {\bibfnamefont {S.~S.}\ \bibnamefont {Malik}}, \ and\
  \bibinfo {author} {\bibfnamefont {A.~M.}\ \bibnamefont {Desai}},\ }\href
  {\doibase 10.1103/PhysRevC.85.065503} {\bibfield  {journal} {\bibinfo
  {journal} {Phys. Rev. C}\ }\textbf {\bibinfo {volume} {85}},\ \bibinfo
  {pages} {065503} (\bibinfo {year} {2012})}\BibitemShut {NoStop}%
\bibitem [{\citenamefont {Arzumanov}\ \emph {et~al.}(2015)\citenamefont
  {Arzumanov}, \citenamefont {Bondarenko}, \citenamefont {Chernyavsky},
  \citenamefont {Geltenbort}, \citenamefont {Morozov}, \citenamefont
  {Nesvizhevsky}, \citenamefont {Panin},\ and\ \citenamefont
  {Strepetov}}]{Arzumanov2015}%
  \BibitemOpen
  \bibfield  {author} {\bibinfo {author} {\bibfnamefont {S.}~\bibnamefont
  {Arzumanov}}, \bibinfo {author} {\bibfnamefont {L.}~\bibnamefont
  {Bondarenko}}, \bibinfo {author} {\bibfnamefont {S.}~\bibnamefont
  {Chernyavsky}}, \bibinfo {author} {\bibfnamefont {P.}~\bibnamefont
  {Geltenbort}}, \bibinfo {author} {\bibfnamefont {V.}~\bibnamefont {Morozov}},
  \bibinfo {author} {\bibfnamefont {V.}~\bibnamefont {Nesvizhevsky}}, \bibinfo
  {author} {\bibfnamefont {Y.}~\bibnamefont {Panin}}, \ and\ \bibinfo {author}
  {\bibfnamefont {A.}~\bibnamefont {Strepetov}},\ }\href {\doibase
  https://doi.org/10.1016/j.physletb.2015.04.021} {\bibfield  {journal}
  {\bibinfo  {journal} {Physics Letters B}\ }\textbf {\bibinfo {volume}
  {745}},\ \bibinfo {pages} {79 } (\bibinfo {year} {2015})}\BibitemShut
  {NoStop}%
\bibitem [{\citenamefont {Ezhov}\ \emph {et~al.}(2018)\citenamefont {Ezhov},
  \citenamefont {Andreev}, \citenamefont {Ban}, \citenamefont {Bazarov},
  \citenamefont {Geltenbort}, \citenamefont {Glushkov}, \citenamefont
  {Knyazkov}, \citenamefont {Kovrizhnykh}, \citenamefont {Krygin},
  \citenamefont {Naviliat-Cuncic},\ and\ \citenamefont {Ryabov}}]{Ezhov2018}%
  \BibitemOpen
  \bibfield  {author} {\bibinfo {author} {\bibfnamefont {V.~F.}\ \bibnamefont
  {Ezhov}}, \bibinfo {author} {\bibfnamefont {A.~Z.}\ \bibnamefont {Andreev}},
  \bibinfo {author} {\bibfnamefont {G.}~\bibnamefont {Ban}}, \bibinfo {author}
  {\bibfnamefont {B.~A.}\ \bibnamefont {Bazarov}}, \bibinfo {author}
  {\bibfnamefont {P.}~\bibnamefont {Geltenbort}}, \bibinfo {author}
  {\bibfnamefont {A.~G.}\ \bibnamefont {Glushkov}}, \bibinfo {author}
  {\bibfnamefont {V.~A.}\ \bibnamefont {Knyazkov}}, \bibinfo {author}
  {\bibfnamefont {N.~A.}\ \bibnamefont {Kovrizhnykh}}, \bibinfo {author}
  {\bibfnamefont {G.~B.}\ \bibnamefont {Krygin}}, \bibinfo {author}
  {\bibfnamefont {O.}~\bibnamefont {Naviliat-Cuncic}}, \ and\ \bibinfo {author}
  {\bibfnamefont {V.~L.}\ \bibnamefont {Ryabov}},\ }\href {\doibase
  10.1134/S0021364018110024} {\bibfield  {journal} {\bibinfo  {journal} {JETP
  Letters}\ }\textbf {\bibinfo {volume} {107}},\ \bibinfo {pages} {671}
  (\bibinfo {year} {2018})}\BibitemShut {NoStop}%
\bibitem [{\citenamefont {Serebrov}\ \emph {et~al.}(2018)\citenamefont
  {Serebrov}, \citenamefont {Kolomensky}, \citenamefont {Fomin}, \citenamefont
  {Krasnoshchekova}, \citenamefont {Vassiljev}, \citenamefont {Prudnikov},
  \citenamefont {Shoka}, \citenamefont {Chechkin}, \citenamefont {Chaikovskiy},
  \citenamefont {Varlamov}, \citenamefont {Ivanov}, \citenamefont {Pirozhkov},
  \citenamefont {Geltenbort}, \citenamefont {Zimmer}, \citenamefont {Jenke},
  \citenamefont {Van~der Grinten},\ and\ \citenamefont
  {Tucker}}]{Serebrov2018}%
  \BibitemOpen
  \bibfield  {author} {\bibinfo {author} {\bibfnamefont {A.~P.}\ \bibnamefont
  {Serebrov}}, \bibinfo {author} {\bibfnamefont {E.~A.}\ \bibnamefont
  {Kolomensky}}, \bibinfo {author} {\bibfnamefont {A.~K.}\ \bibnamefont
  {Fomin}}, \bibinfo {author} {\bibfnamefont {I.~A.}\ \bibnamefont
  {Krasnoshchekova}}, \bibinfo {author} {\bibfnamefont {A.~V.}\ \bibnamefont
  {Vassiljev}}, \bibinfo {author} {\bibfnamefont {D.~M.}\ \bibnamefont
  {Prudnikov}}, \bibinfo {author} {\bibfnamefont {I.~V.}\ \bibnamefont
  {Shoka}}, \bibinfo {author} {\bibfnamefont {A.~V.}\ \bibnamefont {Chechkin}},
  \bibinfo {author} {\bibfnamefont {M.~E.}\ \bibnamefont {Chaikovskiy}},
  \bibinfo {author} {\bibfnamefont {V.~E.}\ \bibnamefont {Varlamov}}, \bibinfo
  {author} {\bibfnamefont {S.~N.}\ \bibnamefont {Ivanov}}, \bibinfo {author}
  {\bibfnamefont {A.~N.}\ \bibnamefont {Pirozhkov}}, \bibinfo {author}
  {\bibfnamefont {P.}~\bibnamefont {Geltenbort}}, \bibinfo {author}
  {\bibfnamefont {O.}~\bibnamefont {Zimmer}}, \bibinfo {author} {\bibfnamefont
  {T.}~\bibnamefont {Jenke}}, \bibinfo {author} {\bibfnamefont
  {M.}~\bibnamefont {Van~der Grinten}}, \ and\ \bibinfo {author} {\bibfnamefont
  {M.}~\bibnamefont {Tucker}},\ }\href {\doibase 10.1103/PhysRevC.97.055503}
  {\bibfield  {journal} {\bibinfo  {journal} {Phys. Rev. C}\ }\textbf {\bibinfo
  {volume} {97}},\ \bibinfo {pages} {055503} (\bibinfo {year}
  {2018})}\BibitemShut {NoStop}%
\bibitem [{\citenamefont {Fornal}\ and\ \citenamefont
  {Grinstein}(2018)}]{Fornal2018}%
  \BibitemOpen
  \bibfield  {author} {\bibinfo {author} {\bibfnamefont {B.}~\bibnamefont
  {Fornal}}\ and\ \bibinfo {author} {\bibfnamefont {B.}~\bibnamefont
  {Grinstein}},\ }\href {\doibase 10.1103/PhysRevLett.120.191801} {\bibfield
  {journal} {\bibinfo  {journal} {Phys. Rev. Lett.}\ }\textbf {\bibinfo
  {volume} {120}},\ \bibinfo {pages} {191801} (\bibinfo {year}
  {2018})}\BibitemShut {NoStop}%
\bibitem [{\citenamefont {Baym}\ \emph {et~al.}(2018)\citenamefont {Baym},
  \citenamefont {Beck}, \citenamefont {Geltenbort},\ and\ \citenamefont
  {Shelton}}]{Baym2018}%
  \BibitemOpen
  \bibfield  {author} {\bibinfo {author} {\bibfnamefont {G.}~\bibnamefont
  {Baym}}, \bibinfo {author} {\bibfnamefont {D.~H.}\ \bibnamefont {Beck}},
  \bibinfo {author} {\bibfnamefont {P.}~\bibnamefont {Geltenbort}}, \ and\
  \bibinfo {author} {\bibfnamefont {J.}~\bibnamefont {Shelton}},\ }\href
  {\doibase 10.1103/PhysRevLett.121.061801} {\bibfield  {journal} {\bibinfo
  {journal} {Phys. Rev. Lett.}\ }\textbf {\bibinfo {volume} {121}},\ \bibinfo
  {pages} {061801} (\bibinfo {year} {2018})}\BibitemShut {NoStop}%
\bibitem [{\citenamefont {McKeen}\ \emph {et~al.}(2018)\citenamefont {McKeen},
  \citenamefont {Nelson}, \citenamefont {Reddy},\ and\ \citenamefont
  {Zhou}}]{McKeen2018}%
  \BibitemOpen
  \bibfield  {author} {\bibinfo {author} {\bibfnamefont {D.}~\bibnamefont
  {McKeen}}, \bibinfo {author} {\bibfnamefont {A.~E.}\ \bibnamefont {Nelson}},
  \bibinfo {author} {\bibfnamefont {S.}~\bibnamefont {Reddy}}, \ and\ \bibinfo
  {author} {\bibfnamefont {D.}~\bibnamefont {Zhou}},\ }\href {\doibase
  10.1103/PhysRevLett.121.061802} {\bibfield  {journal} {\bibinfo  {journal}
  {Phys. Rev. Lett.}\ }\textbf {\bibinfo {volume} {121}},\ \bibinfo {pages}
  {061802} (\bibinfo {year} {2018})}\BibitemShut {NoStop}%
\bibitem [{\citenamefont {Ellis}\ \emph {et~al.}(2018)\citenamefont {Ellis},
  \citenamefont {H\"utsi}, \citenamefont {Kannike}, \citenamefont {Marzola},
  \citenamefont {Raidal},\ and\ \citenamefont {Vaskonen}}]{Ellis2018}%
  \BibitemOpen
  \bibfield  {author} {\bibinfo {author} {\bibfnamefont {J.}~\bibnamefont
  {Ellis}}, \bibinfo {author} {\bibfnamefont {G.}~\bibnamefont {H\"utsi}},
  \bibinfo {author} {\bibfnamefont {K.}~\bibnamefont {Kannike}}, \bibinfo
  {author} {\bibfnamefont {L.}~\bibnamefont {Marzola}}, \bibinfo {author}
  {\bibfnamefont {M.}~\bibnamefont {Raidal}}, \ and\ \bibinfo {author}
  {\bibfnamefont {V.}~\bibnamefont {Vaskonen}},\ }\href {\doibase
  10.1103/PhysRevD.97.123007} {\bibfield  {journal} {\bibinfo  {journal} {Phys.
  Rev. D}\ }\textbf {\bibinfo {volume} {97}},\ \bibinfo {pages} {123007}
  (\bibinfo {year} {2018})}\BibitemShut {NoStop}%
\bibitem [{\citenamefont {McKeen}\ \emph {et~al.}(2021)\citenamefont {McKeen},
  \citenamefont {Pospelov},\ and\ \citenamefont {Raj}}]{McKeen2021A}%
  \BibitemOpen
  \bibfield  {author} {\bibinfo {author} {\bibfnamefont {D.}~\bibnamefont
  {McKeen}}, \bibinfo {author} {\bibfnamefont {M.}~\bibnamefont {Pospelov}}, \
  and\ \bibinfo {author} {\bibfnamefont {N.}~\bibnamefont {Raj}},\ }\href
  {\doibase 10.1103/PhysRevD.103.115002} {\bibfield  {journal} {\bibinfo
  {journal} {Phys. Rev. D}\ }\textbf {\bibinfo {volume} {103}},\ \bibinfo
  {pages} {115002} (\bibinfo {year} {2021})}\BibitemShut {NoStop}%
\bibitem [{\citenamefont {Sun}\ \emph {et~al.}(2018)\citenamefont {Sun},
  \citenamefont {Adamek}, \citenamefont {Allgeier}, \citenamefont {Blatnik},
  \citenamefont {Bowles}, \citenamefont {Broussard}, \citenamefont {Brown},
  \citenamefont {Carr}, \citenamefont {Clayton}, \citenamefont {Cude-Woods},
  \citenamefont {Currie}, \citenamefont {Dees}, \citenamefont {Ding},
  \citenamefont {Filippone}, \citenamefont {Garc\'{\i}a}, \citenamefont
  {Geltenbort}, \citenamefont {Hasan}, \citenamefont {Hickerson}, \citenamefont
  {Hoagland}, \citenamefont {Hong}, \citenamefont {Hogan}, \citenamefont
  {Holley}, \citenamefont {Ito}, \citenamefont {Knecht}, \citenamefont {Liu},
  \citenamefont {Liu}, \citenamefont {Makela}, \citenamefont {Mammei},
  \citenamefont {Martin}, \citenamefont {Melconian}, \citenamefont
  {Mendenhall}, \citenamefont {Moore}, \citenamefont {Morris}, \citenamefont
  {Nepal}, \citenamefont {Nouri}, \citenamefont {Pattie}, \citenamefont
  {P\'erez~Galv\'an}, \citenamefont {Phillips}, \citenamefont {Picker},
  \citenamefont {Pitt}, \citenamefont {Plaster}, \citenamefont {Ramsey},
  \citenamefont {Rios}, \citenamefont {Salvat}, \citenamefont {Saunders},
  \citenamefont {Sondheim}, \citenamefont {Sjue}, \citenamefont {Slutsky},
  \citenamefont {Swank}, \citenamefont {Swift}, \citenamefont {Tatar},
  \citenamefont {Vogelaar}, \citenamefont {VornDick}, \citenamefont {Wang},
  \citenamefont {Wei}, \citenamefont {Wexler}, \citenamefont {Womack},
  \citenamefont {Wrede}, \citenamefont {Young},\ and\ \citenamefont
  {Zeck}}]{Sun2018}%
  \BibitemOpen
  \bibfield  {author} {\bibinfo {author} {\bibfnamefont {X.}~\bibnamefont
  {Sun}}, \bibinfo {author} {\bibfnamefont {E.}~\bibnamefont {Adamek}},
  \bibinfo {author} {\bibfnamefont {B.}~\bibnamefont {Allgeier}}, \bibinfo
  {author} {\bibfnamefont {M.}~\bibnamefont {Blatnik}}, \bibinfo {author}
  {\bibfnamefont {T.~J.}\ \bibnamefont {Bowles}}, \bibinfo {author}
  {\bibfnamefont {L.~J.}\ \bibnamefont {Broussard}}, \bibinfo {author}
  {\bibfnamefont {M.~A.-P.}\ \bibnamefont {Brown}}, \bibinfo {author}
  {\bibfnamefont {R.}~\bibnamefont {Carr}}, \bibinfo {author} {\bibfnamefont
  {S.}~\bibnamefont {Clayton}}, \bibinfo {author} {\bibfnamefont
  {C.}~\bibnamefont {Cude-Woods}}, \bibinfo {author} {\bibfnamefont
  {S.}~\bibnamefont {Currie}}, \bibinfo {author} {\bibfnamefont {E.~B.}\
  \bibnamefont {Dees}}, \bibinfo {author} {\bibfnamefont {X.}~\bibnamefont
  {Ding}}, \bibinfo {author} {\bibfnamefont {B.~W.}\ \bibnamefont {Filippone}},
  \bibinfo {author} {\bibfnamefont {A.}~\bibnamefont {Garc\'{\i}a}}, \bibinfo
  {author} {\bibfnamefont {P.}~\bibnamefont {Geltenbort}}, \bibinfo {author}
  {\bibfnamefont {S.}~\bibnamefont {Hasan}}, \bibinfo {author} {\bibfnamefont
  {K.~P.}\ \bibnamefont {Hickerson}}, \bibinfo {author} {\bibfnamefont
  {J.}~\bibnamefont {Hoagland}}, \bibinfo {author} {\bibfnamefont
  {R.}~\bibnamefont {Hong}}, \bibinfo {author} {\bibfnamefont {G.~E.}\
  \bibnamefont {Hogan}}, \bibinfo {author} {\bibfnamefont {A.~T.}\ \bibnamefont
  {Holley}}, \bibinfo {author} {\bibfnamefont {T.~M.}\ \bibnamefont {Ito}},
  \bibinfo {author} {\bibfnamefont {A.}~\bibnamefont {Knecht}}, \bibinfo
  {author} {\bibfnamefont {C.-Y.}\ \bibnamefont {Liu}}, \bibinfo {author}
  {\bibfnamefont {J.}~\bibnamefont {Liu}}, \bibinfo {author} {\bibfnamefont
  {M.}~\bibnamefont {Makela}}, \bibinfo {author} {\bibfnamefont
  {R.}~\bibnamefont {Mammei}}, \bibinfo {author} {\bibfnamefont {J.~W.}\
  \bibnamefont {Martin}}, \bibinfo {author} {\bibfnamefont {D.}~\bibnamefont
  {Melconian}}, \bibinfo {author} {\bibfnamefont {M.~P.}\ \bibnamefont
  {Mendenhall}}, \bibinfo {author} {\bibfnamefont {S.~D.}\ \bibnamefont
  {Moore}}, \bibinfo {author} {\bibfnamefont {C.~L.}\ \bibnamefont {Morris}},
  \bibinfo {author} {\bibfnamefont {S.}~\bibnamefont {Nepal}}, \bibinfo
  {author} {\bibfnamefont {N.}~\bibnamefont {Nouri}}, \bibinfo {author}
  {\bibfnamefont {R.~W.}\ \bibnamefont {Pattie}}, \bibinfo {author}
  {\bibfnamefont {A.}~\bibnamefont {P\'erez~Galv\'an}}, \bibinfo {author}
  {\bibfnamefont {D.~G.}\ \bibnamefont {Phillips}}, \bibinfo {author}
  {\bibfnamefont {R.}~\bibnamefont {Picker}}, \bibinfo {author} {\bibfnamefont
  {M.~L.}\ \bibnamefont {Pitt}}, \bibinfo {author} {\bibfnamefont
  {B.}~\bibnamefont {Plaster}}, \bibinfo {author} {\bibfnamefont {J.~C.}\
  \bibnamefont {Ramsey}}, \bibinfo {author} {\bibfnamefont {R.}~\bibnamefont
  {Rios}}, \bibinfo {author} {\bibfnamefont {D.~J.}\ \bibnamefont {Salvat}},
  \bibinfo {author} {\bibfnamefont {A.}~\bibnamefont {Saunders}}, \bibinfo
  {author} {\bibfnamefont {W.}~\bibnamefont {Sondheim}}, \bibinfo {author}
  {\bibfnamefont {S.}~\bibnamefont {Sjue}}, \bibinfo {author} {\bibfnamefont
  {S.}~\bibnamefont {Slutsky}}, \bibinfo {author} {\bibfnamefont
  {C.}~\bibnamefont {Swank}}, \bibinfo {author} {\bibfnamefont
  {G.}~\bibnamefont {Swift}}, \bibinfo {author} {\bibfnamefont
  {E.}~\bibnamefont {Tatar}}, \bibinfo {author} {\bibfnamefont {R.~B.}\
  \bibnamefont {Vogelaar}}, \bibinfo {author} {\bibfnamefont {B.}~\bibnamefont
  {VornDick}}, \bibinfo {author} {\bibfnamefont {Z.}~\bibnamefont {Wang}},
  \bibinfo {author} {\bibfnamefont {W.}~\bibnamefont {Wei}}, \bibinfo {author}
  {\bibfnamefont {J.}~\bibnamefont {Wexler}}, \bibinfo {author} {\bibfnamefont
  {T.}~\bibnamefont {Womack}}, \bibinfo {author} {\bibfnamefont
  {C.}~\bibnamefont {Wrede}}, \bibinfo {author} {\bibfnamefont {A.~R.}\
  \bibnamefont {Young}}, \ and\ \bibinfo {author} {\bibfnamefont {B.~A.}\
  \bibnamefont {Zeck}} (\bibinfo {collaboration} {UCNA Collaboration}),\ }\href
  {\doibase 10.1103/PhysRevC.97.052501} {\bibfield  {journal} {\bibinfo
  {journal} {Phys. Rev. C}\ }\textbf {\bibinfo {volume} {97}},\ \bibinfo
  {pages} {052501} (\bibinfo {year} {2018})}\BibitemShut {NoStop}%
\bibitem [{\citenamefont {Tang}\ \emph {et~al.}(2018)\citenamefont {Tang},
  \citenamefont {Blatnik}, \citenamefont {Broussard}, \citenamefont {Choi},
  \citenamefont {Clayton}, \citenamefont {Cude-Woods}, \citenamefont {Currie},
  \citenamefont {Fellers}, \citenamefont {Fries}, \citenamefont {Geltenbort},
  \citenamefont {Gonzalez}, \citenamefont {Hickerson}, \citenamefont {Ito},
  \citenamefont {Liu}, \citenamefont {MacDonald}, \citenamefont {Makela},
  \citenamefont {Morris}, \citenamefont {O'Shaughnessy}, \citenamefont
  {Pattie}, \citenamefont {Plaster}, \citenamefont {Salvat}, \citenamefont
  {Saunders}, \citenamefont {Wang}, \citenamefont {Young},\ and\ \citenamefont
  {Zeck}}]{Tang2018}%
  \BibitemOpen
  \bibfield  {author} {\bibinfo {author} {\bibfnamefont {Z.}~\bibnamefont
  {Tang}}, \bibinfo {author} {\bibfnamefont {M.}~\bibnamefont {Blatnik}},
  \bibinfo {author} {\bibfnamefont {L.~J.}\ \bibnamefont {Broussard}}, \bibinfo
  {author} {\bibfnamefont {J.~H.}\ \bibnamefont {Choi}}, \bibinfo {author}
  {\bibfnamefont {S.~M.}\ \bibnamefont {Clayton}}, \bibinfo {author}
  {\bibfnamefont {C.}~\bibnamefont {Cude-Woods}}, \bibinfo {author}
  {\bibfnamefont {S.}~\bibnamefont {Currie}}, \bibinfo {author} {\bibfnamefont
  {D.~E.}\ \bibnamefont {Fellers}}, \bibinfo {author} {\bibfnamefont {E.~M.}\
  \bibnamefont {Fries}}, \bibinfo {author} {\bibfnamefont {P.}~\bibnamefont
  {Geltenbort}}, \bibinfo {author} {\bibfnamefont {F.}~\bibnamefont
  {Gonzalez}}, \bibinfo {author} {\bibfnamefont {K.~P.}\ \bibnamefont
  {Hickerson}}, \bibinfo {author} {\bibfnamefont {T.~M.}\ \bibnamefont {Ito}},
  \bibinfo {author} {\bibfnamefont {C.-Y.}\ \bibnamefont {Liu}}, \bibinfo
  {author} {\bibfnamefont {S.~W.~T.}\ \bibnamefont {MacDonald}}, \bibinfo
  {author} {\bibfnamefont {M.}~\bibnamefont {Makela}}, \bibinfo {author}
  {\bibfnamefont {C.~L.}\ \bibnamefont {Morris}}, \bibinfo {author}
  {\bibfnamefont {C.~M.}\ \bibnamefont {O'Shaughnessy}}, \bibinfo {author}
  {\bibfnamefont {R.~W.}\ \bibnamefont {Pattie}}, \bibinfo {author}
  {\bibfnamefont {B.}~\bibnamefont {Plaster}}, \bibinfo {author} {\bibfnamefont
  {D.~J.}\ \bibnamefont {Salvat}}, \bibinfo {author} {\bibfnamefont
  {A.}~\bibnamefont {Saunders}}, \bibinfo {author} {\bibfnamefont
  {Z.}~\bibnamefont {Wang}}, \bibinfo {author} {\bibfnamefont {A.~R.}\
  \bibnamefont {Young}}, \ and\ \bibinfo {author} {\bibfnamefont {B.~A.}\
  \bibnamefont {Zeck}},\ }\href {\doibase 10.1103/PhysRevLett.121.022505}
  {\bibfield  {journal} {\bibinfo  {journal} {Phys. Rev. Lett.}\ }\textbf
  {\bibinfo {volume} {121}},\ \bibinfo {pages} {022505} (\bibinfo {year}
  {2018})}\BibitemShut {NoStop}%
\bibitem [{\citenamefont {Wilson}\ \emph {et~al.}(2020)\citenamefont {Wilson},
  \citenamefont {Lawrence}, \citenamefont {Peplowski}, \citenamefont {Eke},\
  and\ \citenamefont {Kegerreis}}]{Wilson2020}%
  \BibitemOpen
  \bibfield  {author} {\bibinfo {author} {\bibfnamefont {J.~T.}\ \bibnamefont
  {Wilson}}, \bibinfo {author} {\bibfnamefont {D.~J.}\ \bibnamefont
  {Lawrence}}, \bibinfo {author} {\bibfnamefont {P.~N.}\ \bibnamefont
  {Peplowski}}, \bibinfo {author} {\bibfnamefont {V.~R.}\ \bibnamefont {Eke}},
  \ and\ \bibinfo {author} {\bibfnamefont {J.~A.}\ \bibnamefont {Kegerreis}},\
  }\href {\doibase 10.1103/PhysRevResearch.2.023316} {\bibfield  {journal}
  {\bibinfo  {journal} {Phys. Rev. Research}\ }\textbf {\bibinfo {volume}
  {2}},\ \bibinfo {pages} {023316} (\bibinfo {year} {2020})}\BibitemShut
  {NoStop}%
\bibitem [{\citenamefont {Lawrence}\ \emph {et~al.}(2021)\citenamefont
  {Lawrence}, \citenamefont {Wilson},\ and\ \citenamefont
  {Peplowski}}]{Lawrence2021}%
  \BibitemOpen
  \bibfield  {author} {\bibinfo {author} {\bibfnamefont {D.~J.}\ \bibnamefont
  {Lawrence}}, \bibinfo {author} {\bibfnamefont {J.~T.}\ \bibnamefont
  {Wilson}}, \ and\ \bibinfo {author} {\bibfnamefont {P.~N.}\ \bibnamefont
  {Peplowski}},\ }\href {\doibase https://doi.org/10.1016/j.nima.2020.164919}
  {\bibfield  {journal} {\bibinfo  {journal} {Nuclear Instruments and Methods
  in Physics Research Section A: Accelerators, Spectrometers, Detectors and
  Associated Equipment}\ }\textbf {\bibinfo {volume} {988}},\ \bibinfo {pages}
  {164919} (\bibinfo {year} {2021})}\BibitemShut {NoStop}%
\bibitem [{\citenamefont {Sumi}\ \emph {et~al.}()\citenamefont {Sumi},
  \citenamefont {Hirota}, \citenamefont {Ichikawa}, \citenamefont {Ino},
  \citenamefont {Iwashita}, \citenamefont {Kajiwara}, \citenamefont {Kato},
  \citenamefont {Kitaguchi}, \citenamefont {Mishima}, \citenamefont {Morikawa},
  \citenamefont {Mogi}, \citenamefont {Oide}, \citenamefont {Okabe},
  \citenamefont {Otono}, \citenamefont {Shima}, \citenamefont {Shimizu},
  \citenamefont {Sugisawa}, \citenamefont {Tanabe}, \citenamefont {Yamashita},
  \citenamefont {Yano},\ and\ \citenamefont {Yoshioka}}]{jparclifetime2019}%
  \BibitemOpen
  \bibfield  {author} {\bibinfo {author} {\bibfnamefont {N.}~\bibnamefont
  {Sumi}}, \bibinfo {author} {\bibfnamefont {K.}~\bibnamefont {Hirota}},
  \bibinfo {author} {\bibfnamefont {G.}~\bibnamefont {Ichikawa}}, \bibinfo
  {author} {\bibfnamefont {T.}~\bibnamefont {Ino}}, \bibinfo {author}
  {\bibfnamefont {Y.}~\bibnamefont {Iwashita}}, \bibinfo {author}
  {\bibfnamefont {S.}~\bibnamefont {Kajiwara}}, \bibinfo {author}
  {\bibfnamefont {Y.}~\bibnamefont {Kato}}, \bibinfo {author} {\bibfnamefont
  {M.}~\bibnamefont {Kitaguchi}}, \bibinfo {author} {\bibfnamefont
  {K.}~\bibnamefont {Mishima}}, \bibinfo {author} {\bibfnamefont
  {K.}~\bibnamefont {Morikawa}}, \bibinfo {author} {\bibfnamefont
  {T.}~\bibnamefont {Mogi}}, \bibinfo {author} {\bibfnamefont {H.}~\bibnamefont
  {Oide}}, \bibinfo {author} {\bibfnamefont {H.}~\bibnamefont {Okabe}},
  \bibinfo {author} {\bibfnamefont {H.}~\bibnamefont {Otono}}, \bibinfo
  {author} {\bibfnamefont {T.}~\bibnamefont {Shima}}, \bibinfo {author}
  {\bibfnamefont {H.~M.}\ \bibnamefont {Shimizu}}, \bibinfo {author}
  {\bibfnamefont {Y.}~\bibnamefont {Sugisawa}}, \bibinfo {author}
  {\bibfnamefont {T.}~\bibnamefont {Tanabe}}, \bibinfo {author} {\bibfnamefont
  {S.}~\bibnamefont {Yamashita}}, \bibinfo {author} {\bibfnamefont
  {K.}~\bibnamefont {Yano}}, \ and\ \bibinfo {author} {\bibfnamefont
  {T.}~\bibnamefont {Yoshioka}},\ }\enquote {\bibinfo {title} {Precise neutron
  lifetime measurement using pulsed neutron beams at j-parc},}\ in\ \href
  {\doibase 10.7566/JPSCP.33.011056} {\emph {\bibinfo {booktitle} {Proceedings
  of the 3rd J-PARC Symposium (J-PARC2019)}}},\ \Eprint
  {http://arxiv.org/abs/https://journals.jps.jp/doi/pdf/10.7566/JPSCP.33.011056}
  {https://journals.jps.jp/doi/pdf/10.7566/JPSCP.33.011056} \BibitemShut
  {NoStop}%
\bibitem [{\citenamefont {Salvat}\ \emph {et~al.}(2014)\citenamefont {Salvat},
  \citenamefont {Adamek}, \citenamefont {Barlow}, \citenamefont {Bowman},
  \citenamefont {Broussard}, \citenamefont {Callahan}, \citenamefont {Clayton},
  \citenamefont {Cude-Woods}, \citenamefont {Currie}, \citenamefont {Dees},
  \citenamefont {Fox}, \citenamefont {Geltenbort}, \citenamefont {Hickerson},
  \citenamefont {Holley}, \citenamefont {Liu}, \citenamefont {Makela},
  \citenamefont {Medina}, \citenamefont {Morley}, \citenamefont {Morris},
  \citenamefont {Penttil\"a}, \citenamefont {Ramsey}, \citenamefont {Saunders},
  \citenamefont {Seestrom}, \citenamefont {Sharapov}, \citenamefont {Sjue},
  \citenamefont {Slaughter}, \citenamefont {Vanderwerp}, \citenamefont
  {VornDick}, \citenamefont {Walstrom}, \citenamefont {Wang}, \citenamefont
  {Womack},\ and\ \citenamefont {Young}}]{Salvat2014}%
  \BibitemOpen
  \bibfield  {author} {\bibinfo {author} {\bibfnamefont {D.~J.}\ \bibnamefont
  {Salvat}}, \bibinfo {author} {\bibfnamefont {E.~R.}\ \bibnamefont {Adamek}},
  \bibinfo {author} {\bibfnamefont {D.}~\bibnamefont {Barlow}}, \bibinfo
  {author} {\bibfnamefont {J.~D.}\ \bibnamefont {Bowman}}, \bibinfo {author}
  {\bibfnamefont {L.~J.}\ \bibnamefont {Broussard}}, \bibinfo {author}
  {\bibfnamefont {N.~B.}\ \bibnamefont {Callahan}}, \bibinfo {author}
  {\bibfnamefont {S.~M.}\ \bibnamefont {Clayton}}, \bibinfo {author}
  {\bibfnamefont {C.}~\bibnamefont {Cude-Woods}}, \bibinfo {author}
  {\bibfnamefont {S.}~\bibnamefont {Currie}}, \bibinfo {author} {\bibfnamefont
  {E.~B.}\ \bibnamefont {Dees}}, \bibinfo {author} {\bibfnamefont
  {W.}~\bibnamefont {Fox}}, \bibinfo {author} {\bibfnamefont {P.}~\bibnamefont
  {Geltenbort}}, \bibinfo {author} {\bibfnamefont {K.~P.}\ \bibnamefont
  {Hickerson}}, \bibinfo {author} {\bibfnamefont {A.~T.}\ \bibnamefont
  {Holley}}, \bibinfo {author} {\bibfnamefont {C.-Y.}\ \bibnamefont {Liu}},
  \bibinfo {author} {\bibfnamefont {M.}~\bibnamefont {Makela}}, \bibinfo
  {author} {\bibfnamefont {J.}~\bibnamefont {Medina}}, \bibinfo {author}
  {\bibfnamefont {D.~J.}\ \bibnamefont {Morley}}, \bibinfo {author}
  {\bibfnamefont {C.~L.}\ \bibnamefont {Morris}}, \bibinfo {author}
  {\bibfnamefont {S.~I.}\ \bibnamefont {Penttil\"a}}, \bibinfo {author}
  {\bibfnamefont {J.}~\bibnamefont {Ramsey}}, \bibinfo {author} {\bibfnamefont
  {A.}~\bibnamefont {Saunders}}, \bibinfo {author} {\bibfnamefont {S.~J.}\
  \bibnamefont {Seestrom}}, \bibinfo {author} {\bibfnamefont {E.~I.}\
  \bibnamefont {Sharapov}}, \bibinfo {author} {\bibfnamefont {S.~K.~L.}\
  \bibnamefont {Sjue}}, \bibinfo {author} {\bibfnamefont {B.~A.}\ \bibnamefont
  {Slaughter}}, \bibinfo {author} {\bibfnamefont {J.}~\bibnamefont
  {Vanderwerp}}, \bibinfo {author} {\bibfnamefont {B.}~\bibnamefont
  {VornDick}}, \bibinfo {author} {\bibfnamefont {P.~L.}\ \bibnamefont
  {Walstrom}}, \bibinfo {author} {\bibfnamefont {Z.}~\bibnamefont {Wang}},
  \bibinfo {author} {\bibfnamefont {T.~L.}\ \bibnamefont {Womack}}, \ and\
  \bibinfo {author} {\bibfnamefont {A.~R.}\ \bibnamefont {Young}},\ }\href
  {\doibase 10.1103/PhysRevC.89.052501} {\bibfield  {journal} {\bibinfo
  {journal} {Phys. Rev. C}\ }\textbf {\bibinfo {volume} {89}},\ \bibinfo
  {pages} {052501} (\bibinfo {year} {2014})}\BibitemShut {NoStop}%
\bibitem [{\citenamefont {Morris}\ \emph {et~al.}(2017)\citenamefont {Morris},
  \citenamefont {Adamek}, \citenamefont {Broussard}, \citenamefont {Callahan},
  \citenamefont {Clayton}, \citenamefont {Cude-Woods}, \citenamefont {Currie},
  \citenamefont {Ding}, \citenamefont {Fox}, \citenamefont {Hickerson},
  \citenamefont {Hoffbauer}, \citenamefont {Holley}, \citenamefont {Komives},
  \citenamefont {Liu}, \citenamefont {Makela}, \citenamefont {Pattie},
  \citenamefont {Ramsey}, \citenamefont {Salvat}, \citenamefont {Saunders},
  \citenamefont {Seestrom}, \citenamefont {Sharapov}, \citenamefont {Sjue},
  \citenamefont {Tang}, \citenamefont {Vanderwerp}, \citenamefont {Vogelaar},
  \citenamefont {Walstrom}, \citenamefont {Wang}, \citenamefont {Wei},
  \citenamefont {Wexler}, \citenamefont {Womack}, \citenamefont {Young},\ and\
  \citenamefont {Zeck}}]{Morris2017}%
  \BibitemOpen
  \bibfield  {author} {\bibinfo {author} {\bibfnamefont {C.~L.}\ \bibnamefont
  {Morris}}, \bibinfo {author} {\bibfnamefont {E.~R.}\ \bibnamefont {Adamek}},
  \bibinfo {author} {\bibfnamefont {L.~J.}\ \bibnamefont {Broussard}}, \bibinfo
  {author} {\bibfnamefont {N.~B.}\ \bibnamefont {Callahan}}, \bibinfo {author}
  {\bibfnamefont {S.~M.}\ \bibnamefont {Clayton}}, \bibinfo {author}
  {\bibfnamefont {C.}~\bibnamefont {Cude-Woods}}, \bibinfo {author}
  {\bibfnamefont {S.~A.}\ \bibnamefont {Currie}}, \bibinfo {author}
  {\bibfnamefont {X.}~\bibnamefont {Ding}}, \bibinfo {author} {\bibfnamefont
  {W.}~\bibnamefont {Fox}}, \bibinfo {author} {\bibfnamefont {K.~P.}\
  \bibnamefont {Hickerson}}, \bibinfo {author} {\bibfnamefont {M.~A.}\
  \bibnamefont {Hoffbauer}}, \bibinfo {author} {\bibfnamefont {A.~T.}\
  \bibnamefont {Holley}}, \bibinfo {author} {\bibfnamefont {A.}~\bibnamefont
  {Komives}}, \bibinfo {author} {\bibfnamefont {C.-Y.}\ \bibnamefont {Liu}},
  \bibinfo {author} {\bibfnamefont {M.}~\bibnamefont {Makela}}, \bibinfo
  {author} {\bibfnamefont {R.~W.}\ \bibnamefont {Pattie}}, \bibinfo {author}
  {\bibfnamefont {J.}~\bibnamefont {Ramsey}}, \bibinfo {author} {\bibfnamefont
  {D.~J.}\ \bibnamefont {Salvat}}, \bibinfo {author} {\bibfnamefont
  {A.}~\bibnamefont {Saunders}}, \bibinfo {author} {\bibfnamefont {S.~J.}\
  \bibnamefont {Seestrom}}, \bibinfo {author} {\bibfnamefont {E.~I.}\
  \bibnamefont {Sharapov}}, \bibinfo {author} {\bibfnamefont {S.~K.}\
  \bibnamefont {Sjue}}, \bibinfo {author} {\bibfnamefont {Z.}~\bibnamefont
  {Tang}}, \bibinfo {author} {\bibfnamefont {J.}~\bibnamefont {Vanderwerp}},
  \bibinfo {author} {\bibfnamefont {B.}~\bibnamefont {Vogelaar}}, \bibinfo
  {author} {\bibfnamefont {P.~L.}\ \bibnamefont {Walstrom}}, \bibinfo {author}
  {\bibfnamefont {Z.}~\bibnamefont {Wang}}, \bibinfo {author} {\bibfnamefont
  {W.}~\bibnamefont {Wei}}, \bibinfo {author} {\bibfnamefont {J.~W.}\
  \bibnamefont {Wexler}}, \bibinfo {author} {\bibfnamefont {T.~L.}\
  \bibnamefont {Womack}}, \bibinfo {author} {\bibfnamefont {A.~R.}\
  \bibnamefont {Young}}, \ and\ \bibinfo {author} {\bibfnamefont {B.~A.}\
  \bibnamefont {Zeck}},\ }\href {\doibase 10.1063/1.4983578} {\bibfield
  {journal} {\bibinfo  {journal} {Review of Scientific Instruments}\ }\textbf
  {\bibinfo {volume} {88}},\ \bibinfo {pages} {053508} (\bibinfo {year}
  {2017})},\ \Eprint {http://arxiv.org/abs/https://doi.org/10.1063/1.4983578}
  {https://doi.org/10.1063/1.4983578} \BibitemShut {NoStop}%
\bibitem [{\citenamefont {Pattie}\ \emph {et~al.}(2018)\citenamefont {Pattie},
  \citenamefont {Callahan}, \citenamefont {Cude-Woods}, \citenamefont {Adamek},
  \citenamefont {Broussard}, \citenamefont {Clayton}, \citenamefont {Currie},
  \citenamefont {Dees}, \citenamefont {Ding}, \citenamefont {Engel},
  \citenamefont {Fellers}, \citenamefont {Fox}, \citenamefont {Geltenbort},
  \citenamefont {Hickerson}, \citenamefont {Hoffbauer}, \citenamefont {Holley},
  \citenamefont {Komives}, \citenamefont {Liu}, \citenamefont {MacDonald},
  \citenamefont {Makela}, \citenamefont {Morris}, \citenamefont {Ortiz},
  \citenamefont {Ramsey}, \citenamefont {Salvat}, \citenamefont {Saunders},
  \citenamefont {Seestrom}, \citenamefont {Sharapov}, \citenamefont {Sjue},
  \citenamefont {Tang}, \citenamefont {Vanderwerp}, \citenamefont {Vogelaar},
  \citenamefont {Walstrom}, \citenamefont {Wang}, \citenamefont {Wei},
  \citenamefont {Weaver}, \citenamefont {Wexler}, \citenamefont {Womack},
  \citenamefont {Young},\ and\ \citenamefont {Zeck}}]{Pattie2018}%
  \BibitemOpen
  \bibfield  {author} {\bibinfo {author} {\bibfnamefont {R.~W.}\ \bibnamefont
  {Pattie}}, \bibinfo {author} {\bibfnamefont {N.~B.}\ \bibnamefont
  {Callahan}}, \bibinfo {author} {\bibfnamefont {C.}~\bibnamefont
  {Cude-Woods}}, \bibinfo {author} {\bibfnamefont {E.~R.}\ \bibnamefont
  {Adamek}}, \bibinfo {author} {\bibfnamefont {L.~J.}\ \bibnamefont
  {Broussard}}, \bibinfo {author} {\bibfnamefont {S.~M.}\ \bibnamefont
  {Clayton}}, \bibinfo {author} {\bibfnamefont {S.~A.}\ \bibnamefont {Currie}},
  \bibinfo {author} {\bibfnamefont {E.~B.}\ \bibnamefont {Dees}}, \bibinfo
  {author} {\bibfnamefont {X.}~\bibnamefont {Ding}}, \bibinfo {author}
  {\bibfnamefont {E.~M.}\ \bibnamefont {Engel}}, \bibinfo {author}
  {\bibfnamefont {D.~E.}\ \bibnamefont {Fellers}}, \bibinfo {author}
  {\bibfnamefont {W.}~\bibnamefont {Fox}}, \bibinfo {author} {\bibfnamefont
  {P.}~\bibnamefont {Geltenbort}}, \bibinfo {author} {\bibfnamefont {K.~P.}\
  \bibnamefont {Hickerson}}, \bibinfo {author} {\bibfnamefont {M.~A.}\
  \bibnamefont {Hoffbauer}}, \bibinfo {author} {\bibfnamefont {A.~T.}\
  \bibnamefont {Holley}}, \bibinfo {author} {\bibfnamefont {A.}~\bibnamefont
  {Komives}}, \bibinfo {author} {\bibfnamefont {C.-Y.}\ \bibnamefont {Liu}},
  \bibinfo {author} {\bibfnamefont {S.~W.~T.}\ \bibnamefont {MacDonald}},
  \bibinfo {author} {\bibfnamefont {M.}~\bibnamefont {Makela}}, \bibinfo
  {author} {\bibfnamefont {C.~L.}\ \bibnamefont {Morris}}, \bibinfo {author}
  {\bibfnamefont {J.~D.}\ \bibnamefont {Ortiz}}, \bibinfo {author}
  {\bibfnamefont {J.}~\bibnamefont {Ramsey}}, \bibinfo {author} {\bibfnamefont
  {D.~J.}\ \bibnamefont {Salvat}}, \bibinfo {author} {\bibfnamefont
  {A.}~\bibnamefont {Saunders}}, \bibinfo {author} {\bibfnamefont {S.~J.}\
  \bibnamefont {Seestrom}}, \bibinfo {author} {\bibfnamefont {E.~I.}\
  \bibnamefont {Sharapov}}, \bibinfo {author} {\bibfnamefont {S.~K.}\
  \bibnamefont {Sjue}}, \bibinfo {author} {\bibfnamefont {Z.}~\bibnamefont
  {Tang}}, \bibinfo {author} {\bibfnamefont {J.}~\bibnamefont {Vanderwerp}},
  \bibinfo {author} {\bibfnamefont {B.}~\bibnamefont {Vogelaar}}, \bibinfo
  {author} {\bibfnamefont {P.~L.}\ \bibnamefont {Walstrom}}, \bibinfo {author}
  {\bibfnamefont {Z.}~\bibnamefont {Wang}}, \bibinfo {author} {\bibfnamefont
  {W.}~\bibnamefont {Wei}}, \bibinfo {author} {\bibfnamefont {H.~L.}\
  \bibnamefont {Weaver}}, \bibinfo {author} {\bibfnamefont {J.~W.}\
  \bibnamefont {Wexler}}, \bibinfo {author} {\bibfnamefont {T.~L.}\
  \bibnamefont {Womack}}, \bibinfo {author} {\bibfnamefont {A.~R.}\
  \bibnamefont {Young}}, \ and\ \bibinfo {author} {\bibfnamefont {B.~A.}\
  \bibnamefont {Zeck}},\ }\href {\doibase 10.1126/science.aan8895} {\bibfield
  {journal} {\bibinfo  {journal} {Science}\ }\textbf {\bibinfo {volume}
  {360}},\ \bibinfo {pages} {627} (\bibinfo {year} {2018})},\ \Eprint
  {http://arxiv.org/abs/https://science.sciencemag.org/content/360/6389/627.full.pdf}
  {https://science.sciencemag.org/content/360/6389/627.full.pdf} \BibitemShut
  {NoStop}%
\bibitem [{\citenamefont {Ito}\ \emph {et~al.}(2018)\citenamefont {Ito},
  \citenamefont {Adamek}, \citenamefont {Callahan}, \citenamefont {Choi},
  \citenamefont {Clayton}, \citenamefont {Cude-Woods}, \citenamefont {Currie},
  \citenamefont {Ding}, \citenamefont {Fellers}, \citenamefont {Geltenbort},
  \citenamefont {Lamoreaux}, \citenamefont {Liu}, \citenamefont {MacDonald},
  \citenamefont {Makela}, \citenamefont {Morris}, \citenamefont {Pattie},
  \citenamefont {Ramsey}, \citenamefont {Salvat}, \citenamefont {Saunders},
  \citenamefont {Sharapov}, \citenamefont {Sjue}, \citenamefont {Sprow},
  \citenamefont {Tang}, \citenamefont {Weaver}, \citenamefont {Wei},\ and\
  \citenamefont {Young}}]{Ito2018}%
  \BibitemOpen
  \bibfield  {author} {\bibinfo {author} {\bibfnamefont {T.~M.}\ \bibnamefont
  {Ito}}, \bibinfo {author} {\bibfnamefont {E.~R.}\ \bibnamefont {Adamek}},
  \bibinfo {author} {\bibfnamefont {N.~B.}\ \bibnamefont {Callahan}}, \bibinfo
  {author} {\bibfnamefont {J.~H.}\ \bibnamefont {Choi}}, \bibinfo {author}
  {\bibfnamefont {S.~M.}\ \bibnamefont {Clayton}}, \bibinfo {author}
  {\bibfnamefont {C.}~\bibnamefont {Cude-Woods}}, \bibinfo {author}
  {\bibfnamefont {S.}~\bibnamefont {Currie}}, \bibinfo {author} {\bibfnamefont
  {X.}~\bibnamefont {Ding}}, \bibinfo {author} {\bibfnamefont {D.~E.}\
  \bibnamefont {Fellers}}, \bibinfo {author} {\bibfnamefont {P.}~\bibnamefont
  {Geltenbort}}, \bibinfo {author} {\bibfnamefont {S.~K.}\ \bibnamefont
  {Lamoreaux}}, \bibinfo {author} {\bibfnamefont {C.-Y.}\ \bibnamefont {Liu}},
  \bibinfo {author} {\bibfnamefont {S.}~\bibnamefont {MacDonald}}, \bibinfo
  {author} {\bibfnamefont {M.}~\bibnamefont {Makela}}, \bibinfo {author}
  {\bibfnamefont {C.~L.}\ \bibnamefont {Morris}}, \bibinfo {author}
  {\bibfnamefont {R.~W.}\ \bibnamefont {Pattie}}, \bibinfo {author}
  {\bibfnamefont {J.~C.}\ \bibnamefont {Ramsey}}, \bibinfo {author}
  {\bibfnamefont {D.~J.}\ \bibnamefont {Salvat}}, \bibinfo {author}
  {\bibfnamefont {A.}~\bibnamefont {Saunders}}, \bibinfo {author}
  {\bibfnamefont {E.~I.}\ \bibnamefont {Sharapov}}, \bibinfo {author}
  {\bibfnamefont {S.}~\bibnamefont {Sjue}}, \bibinfo {author} {\bibfnamefont
  {A.~P.}\ \bibnamefont {Sprow}}, \bibinfo {author} {\bibfnamefont
  {Z.}~\bibnamefont {Tang}}, \bibinfo {author} {\bibfnamefont {H.~L.}\
  \bibnamefont {Weaver}}, \bibinfo {author} {\bibfnamefont {W.}~\bibnamefont
  {Wei}}, \ and\ \bibinfo {author} {\bibfnamefont {A.~R.}\ \bibnamefont
  {Young}},\ }\href {\doibase 10.1103/PhysRevC.97.012501} {\bibfield  {journal}
  {\bibinfo  {journal} {Phys. Rev. C}\ }\textbf {\bibinfo {volume} {97}},\
  \bibinfo {pages} {012501} (\bibinfo {year} {2018})}\BibitemShut {NoStop}%
\bibitem [{\citenamefont {Holley}\ \emph {et~al.}(2012)\citenamefont {Holley},
  \citenamefont {Broussard}, \citenamefont {Davis}, \citenamefont {Hickerson},
  \citenamefont {Ito}, \citenamefont {Liu}, \citenamefont {Lyles},
  \citenamefont {Makela}, \citenamefont {Mammei}, \citenamefont {Mendenhall},
  \citenamefont {Morris}, \citenamefont {Mortensen}, \citenamefont {Pattie},
  \citenamefont {Rios}, \citenamefont {Saunders},\ and\ \citenamefont
  {Young}}]{Holley2012}%
  \BibitemOpen
  \bibfield  {author} {\bibinfo {author} {\bibfnamefont {A.~T.}\ \bibnamefont
  {Holley}}, \bibinfo {author} {\bibfnamefont {L.~J.}\ \bibnamefont
  {Broussard}}, \bibinfo {author} {\bibfnamefont {J.~L.}\ \bibnamefont
  {Davis}}, \bibinfo {author} {\bibfnamefont {K.}~\bibnamefont {Hickerson}},
  \bibinfo {author} {\bibfnamefont {T.~M.}\ \bibnamefont {Ito}}, \bibinfo
  {author} {\bibfnamefont {C.-Y.}\ \bibnamefont {Liu}}, \bibinfo {author}
  {\bibfnamefont {J.~T.~M.}\ \bibnamefont {Lyles}}, \bibinfo {author}
  {\bibfnamefont {M.}~\bibnamefont {Makela}}, \bibinfo {author} {\bibfnamefont
  {R.~R.}\ \bibnamefont {Mammei}}, \bibinfo {author} {\bibfnamefont {M.~P.}\
  \bibnamefont {Mendenhall}}, \bibinfo {author} {\bibfnamefont {C.~L.}\
  \bibnamefont {Morris}}, \bibinfo {author} {\bibfnamefont {R.}~\bibnamefont
  {Mortensen}}, \bibinfo {author} {\bibfnamefont {R.~W.}\ \bibnamefont
  {Pattie}}, \bibinfo {author} {\bibfnamefont {R.}~\bibnamefont {Rios}},
  \bibinfo {author} {\bibfnamefont {A.}~\bibnamefont {Saunders}}, \ and\
  \bibinfo {author} {\bibfnamefont {A.~R.}\ \bibnamefont {Young}},\ }\href
  {\doibase 10.1063/1.4732822} {\bibfield  {journal} {\bibinfo  {journal}
  {Review of Scientific Instruments}\ }\textbf {\bibinfo {volume} {83}},\
  \bibinfo {pages} {073505} (\bibinfo {year} {2012})},\ \Eprint
  {http://arxiv.org/abs/https://doi.org/10.1063/1.4732822}
  {https://doi.org/10.1063/1.4732822} \BibitemShut {NoStop}%
\bibitem [{\citenamefont {Walstrom}\ \emph {et~al.}(2009)\citenamefont
  {Walstrom}, \citenamefont {Bowman}, \citenamefont {Penttila}, \citenamefont
  {Morris},\ and\ \citenamefont {Saunders}}]{Walstrom2009}%
  \BibitemOpen
  \bibfield  {author} {\bibinfo {author} {\bibfnamefont {P.}~\bibnamefont
  {Walstrom}}, \bibinfo {author} {\bibfnamefont {J.}~\bibnamefont {Bowman}},
  \bibinfo {author} {\bibfnamefont {S.}~\bibnamefont {Penttila}}, \bibinfo
  {author} {\bibfnamefont {C.}~\bibnamefont {Morris}}, \ and\ \bibinfo {author}
  {\bibfnamefont {A.}~\bibnamefont {Saunders}},\ }\href {\doibase
  https://doi.org/10.1016/j.nima.2008.11.010} {\bibfield  {journal} {\bibinfo
  {journal} {Nuclear Instruments and Methods in Physics Research Section A:
  Accelerators, Spectrometers, Detectors and Associated Equipment}\ }\textbf
  {\bibinfo {volume} {599}},\ \bibinfo {pages} {82 } (\bibinfo {year}
  {2009})}\BibitemShut {NoStop}%
\bibitem [{\citenamefont {Picker}\ \emph {et~al.}(2009)\citenamefont {Picker},
  \citenamefont {Altarev}, \citenamefont {Amos}, \citenamefont {Franke},
  \citenamefont {Geltenbort}, \citenamefont {Gutsmiedl}, \citenamefont
  {Hartmann}, \citenamefont {Mann}, \citenamefont {Materne}, \citenamefont
  {Müller}, \citenamefont {Paul}, \citenamefont {Stoepler},\ and\
  \citenamefont {Wirth}}]{Picker2009}%
  \BibitemOpen
  \bibfield  {author} {\bibinfo {author} {\bibfnamefont {R.}~\bibnamefont
  {Picker}}, \bibinfo {author} {\bibfnamefont {I.}~\bibnamefont {Altarev}},
  \bibinfo {author} {\bibfnamefont {P.}~\bibnamefont {Amos}}, \bibinfo {author}
  {\bibfnamefont {B.}~\bibnamefont {Franke}}, \bibinfo {author} {\bibfnamefont
  {P.}~\bibnamefont {Geltenbort}}, \bibinfo {author} {\bibfnamefont
  {E.}~\bibnamefont {Gutsmiedl}}, \bibinfo {author} {\bibfnamefont
  {F.}~\bibnamefont {Hartmann}}, \bibinfo {author} {\bibfnamefont
  {A.}~\bibnamefont {Mann}}, \bibinfo {author} {\bibfnamefont {S.}~\bibnamefont
  {Materne}}, \bibinfo {author} {\bibfnamefont {A.}~\bibnamefont {Müller}},
  \bibinfo {author} {\bibfnamefont {S.}~\bibnamefont {Paul}}, \bibinfo {author}
  {\bibfnamefont {R.}~\bibnamefont {Stoepler}}, \ and\ \bibinfo {author}
  {\bibfnamefont {H.-F.}\ \bibnamefont {Wirth}},\ }\href {\doibase
  https://doi.org/10.1016/j.nima.2009.07.089} {\bibfield  {journal} {\bibinfo
  {journal} {Nuclear Instruments and Methods in Physics Research Section A:
  Accelerators, Spectrometers, Detectors and Associated Equipment}\ }\textbf
  {\bibinfo {volume} {611}},\ \bibinfo {pages} {297 } (\bibinfo {year}
  {2009})},\ \bibinfo {note} {particle Physics with Slow Neutrons}\BibitemShut
  {NoStop}%
\bibitem [{\citenamefont {Wang}\ \emph {et~al.}(2015)\citenamefont {Wang},
  \citenamefont {Hoffbauer}, \citenamefont {Morris}, \citenamefont {Callahan},
  \citenamefont {Adamek}, \citenamefont {Bacon}, \citenamefont {Blatnik},
  \citenamefont {Brandt}, \citenamefont {Broussard}, \citenamefont {Clayton},
  \citenamefont {Cude-Woods}, \citenamefont {Currie}, \citenamefont {Dees},
  \citenamefont {Ding}, \citenamefont {Gao}, \citenamefont {Gray},
  \citenamefont {Hickerson}, \citenamefont {Holley}, \citenamefont {Ito},
  \citenamefont {Liu}, \citenamefont {Makela}, \citenamefont {Ramsey},
  \citenamefont {Pattie}, \citenamefont {Salvat}, \citenamefont {Saunders},
  \citenamefont {Schmidt}, \citenamefont {Schulze}, \citenamefont {Seestrom},
  \citenamefont {Sharapov}, \citenamefont {Sprow}, \citenamefont {Tang},
  \citenamefont {Wei}, \citenamefont {Wexler}, \citenamefont {Womack},
  \citenamefont {Young},\ and\ \citenamefont {Zeck}}]{Wang2015}%
  \BibitemOpen
  \bibfield  {author} {\bibinfo {author} {\bibfnamefont {Z.}~\bibnamefont
  {Wang}}, \bibinfo {author} {\bibfnamefont {M.}~\bibnamefont {Hoffbauer}},
  \bibinfo {author} {\bibfnamefont {C.}~\bibnamefont {Morris}}, \bibinfo
  {author} {\bibfnamefont {N.}~\bibnamefont {Callahan}}, \bibinfo {author}
  {\bibfnamefont {E.}~\bibnamefont {Adamek}}, \bibinfo {author} {\bibfnamefont
  {J.}~\bibnamefont {Bacon}}, \bibinfo {author} {\bibfnamefont
  {M.}~\bibnamefont {Blatnik}}, \bibinfo {author} {\bibfnamefont
  {A.}~\bibnamefont {Brandt}}, \bibinfo {author} {\bibfnamefont
  {L.}~\bibnamefont {Broussard}}, \bibinfo {author} {\bibfnamefont
  {S.}~\bibnamefont {Clayton}}, \bibinfo {author} {\bibfnamefont
  {C.}~\bibnamefont {Cude-Woods}}, \bibinfo {author} {\bibfnamefont
  {S.}~\bibnamefont {Currie}}, \bibinfo {author} {\bibfnamefont
  {E.}~\bibnamefont {Dees}}, \bibinfo {author} {\bibfnamefont {X.}~\bibnamefont
  {Ding}}, \bibinfo {author} {\bibfnamefont {J.}~\bibnamefont {Gao}}, \bibinfo
  {author} {\bibfnamefont {F.}~\bibnamefont {Gray}}, \bibinfo {author}
  {\bibfnamefont {K.}~\bibnamefont {Hickerson}}, \bibinfo {author}
  {\bibfnamefont {A.}~\bibnamefont {Holley}}, \bibinfo {author} {\bibfnamefont
  {T.}~\bibnamefont {Ito}}, \bibinfo {author} {\bibfnamefont {C.-Y.}\
  \bibnamefont {Liu}}, \bibinfo {author} {\bibfnamefont {M.}~\bibnamefont
  {Makela}}, \bibinfo {author} {\bibfnamefont {J.}~\bibnamefont {Ramsey}},
  \bibinfo {author} {\bibfnamefont {R.}~\bibnamefont {Pattie}}, \bibinfo
  {author} {\bibfnamefont {D.}~\bibnamefont {Salvat}}, \bibinfo {author}
  {\bibfnamefont {A.}~\bibnamefont {Saunders}}, \bibinfo {author}
  {\bibfnamefont {D.}~\bibnamefont {Schmidt}}, \bibinfo {author} {\bibfnamefont
  {R.}~\bibnamefont {Schulze}}, \bibinfo {author} {\bibfnamefont
  {S.}~\bibnamefont {Seestrom}}, \bibinfo {author} {\bibfnamefont
  {E.}~\bibnamefont {Sharapov}}, \bibinfo {author} {\bibfnamefont
  {A.}~\bibnamefont {Sprow}}, \bibinfo {author} {\bibfnamefont
  {Z.}~\bibnamefont {Tang}}, \bibinfo {author} {\bibfnamefont {W.}~\bibnamefont
  {Wei}}, \bibinfo {author} {\bibfnamefont {J.}~\bibnamefont {Wexler}},
  \bibinfo {author} {\bibfnamefont {T.}~\bibnamefont {Womack}}, \bibinfo
  {author} {\bibfnamefont {A.}~\bibnamefont {Young}}, \ and\ \bibinfo {author}
  {\bibfnamefont {B.}~\bibnamefont {Zeck}},\ }\href {\doibase
  https://doi.org/10.1016/j.nima.2015.07.010} {\bibfield  {journal} {\bibinfo
  {journal} {Nuclear Instruments and Methods in Physics Research Section A:
  Accelerators, Spectrometers, Detectors and Associated Equipment}\ }\textbf
  {\bibinfo {volume} {798}},\ \bibinfo {pages} {30 } (\bibinfo {year}
  {2015})}\BibitemShut {NoStop}%
\bibitem [{\citenamefont {Anghel}\ \emph {et~al.}(2018)\citenamefont {Anghel},
  \citenamefont {Bailey}, \citenamefont {Bison}, \citenamefont {Blau},
  \citenamefont {Broussard}, \citenamefont {Clayton}, \citenamefont
  {Cude-Woods}, \citenamefont {Daum}, \citenamefont {Hawari}, \citenamefont
  {Hild}, \citenamefont {Huffman}, \citenamefont {Ito}, \citenamefont {Kirch},
  \citenamefont {Korobkina}, \citenamefont {Lauss}, \citenamefont {Leung},
  \citenamefont {Lutz}, \citenamefont {Makela}, \citenamefont {Medlin},
  \citenamefont {Morris}, \citenamefont {Pattie}, \citenamefont {Ries},
  \citenamefont {Saunders}, \citenamefont {Schmidt-Wellenburg}, \citenamefont
  {Talanov}, \citenamefont {Young}, \citenamefont {Wehring}, \citenamefont
  {White}, \citenamefont {Wohlmuther},\ and\ \citenamefont
  {Zsigmond}}]{Anghel2018}%
  \BibitemOpen
  \bibfield  {author} {\bibinfo {author} {\bibfnamefont {A.}~\bibnamefont
  {Anghel}}, \bibinfo {author} {\bibfnamefont {T.~L.}\ \bibnamefont {Bailey}},
  \bibinfo {author} {\bibfnamefont {G.}~\bibnamefont {Bison}}, \bibinfo
  {author} {\bibfnamefont {B.}~\bibnamefont {Blau}}, \bibinfo {author}
  {\bibfnamefont {L.~J.}\ \bibnamefont {Broussard}}, \bibinfo {author}
  {\bibfnamefont {S.~M.}\ \bibnamefont {Clayton}}, \bibinfo {author}
  {\bibfnamefont {C.}~\bibnamefont {Cude-Woods}}, \bibinfo {author}
  {\bibfnamefont {M.}~\bibnamefont {Daum}}, \bibinfo {author} {\bibfnamefont
  {A.}~\bibnamefont {Hawari}}, \bibinfo {author} {\bibfnamefont
  {N.}~\bibnamefont {Hild}}, \bibinfo {author} {\bibfnamefont {P.}~\bibnamefont
  {Huffman}}, \bibinfo {author} {\bibfnamefont {T.~M.}\ \bibnamefont {Ito}},
  \bibinfo {author} {\bibfnamefont {K.}~\bibnamefont {Kirch}}, \bibinfo
  {author} {\bibfnamefont {E.}~\bibnamefont {Korobkina}}, \bibinfo {author}
  {\bibfnamefont {B.}~\bibnamefont {Lauss}}, \bibinfo {author} {\bibfnamefont
  {K.}~\bibnamefont {Leung}}, \bibinfo {author} {\bibfnamefont {E.~M.}\
  \bibnamefont {Lutz}}, \bibinfo {author} {\bibfnamefont {M.}~\bibnamefont
  {Makela}}, \bibinfo {author} {\bibfnamefont {G.}~\bibnamefont {Medlin}},
  \bibinfo {author} {\bibfnamefont {C.~L.}\ \bibnamefont {Morris}}, \bibinfo
  {author} {\bibfnamefont {R.~W.}\ \bibnamefont {Pattie}}, \bibinfo {author}
  {\bibfnamefont {D.}~\bibnamefont {Ries}}, \bibinfo {author} {\bibfnamefont
  {A.}~\bibnamefont {Saunders}}, \bibinfo {author} {\bibfnamefont
  {P.}~\bibnamefont {Schmidt-Wellenburg}}, \bibinfo {author} {\bibfnamefont
  {V.}~\bibnamefont {Talanov}}, \bibinfo {author} {\bibfnamefont {A.~R.}\
  \bibnamefont {Young}}, \bibinfo {author} {\bibfnamefont {B.}~\bibnamefont
  {Wehring}}, \bibinfo {author} {\bibfnamefont {C.}~\bibnamefont {White}},
  \bibinfo {author} {\bibfnamefont {M.}~\bibnamefont {Wohlmuther}}, \ and\
  \bibinfo {author} {\bibfnamefont {G.}~\bibnamefont {Zsigmond}},\ }\href
  {\doibase 10.1140/epja/i2018-12594-2} {\bibfield  {journal} {\bibinfo
  {journal} {The European Physical Journal A}\ }\textbf {\bibinfo {volume}
  {54}} (\bibinfo {year} {2018}),\ 10.1140/epja/i2018-12594-2}\BibitemShut
  {NoStop}%
\bibitem [{\citenamefont {Abdi}\ and\ \citenamefont
  {Williams}(2010)}]{principlecomponents}%
  \BibitemOpen
  \bibfield  {author} {\bibinfo {author} {\bibfnamefont {H.}~\bibnamefont
  {Abdi}}\ and\ \bibinfo {author} {\bibfnamefont {L.~J.}\ \bibnamefont
  {Williams}},\ }\href {\doibase https://doi.org/10.1002/wics.101} {\bibfield
  {journal} {\bibinfo  {journal} {WIREs Computational Statistics}\ }\textbf
  {\bibinfo {volume} {2}},\ \bibinfo {pages} {433} (\bibinfo {year} {2010})},\
  \Eprint
  {http://arxiv.org/abs/https://onlinelibrary.wiley.com/doi/pdf/10.1002/wics.101}
  {https://onlinelibrary.wiley.com/doi/pdf/10.1002/wics.101} \BibitemShut
  {NoStop}%
\bibitem [{\citenamefont {Foreman-Mackey}\ \emph {et~al.}(2013)\citenamefont
  {Foreman-Mackey}, \citenamefont {Hogg}, \citenamefont {Lang},\ and\
  \citenamefont {Goodman}}]{ForemanMackey2013}%
  \BibitemOpen
  \bibfield  {author} {\bibinfo {author} {\bibfnamefont {D.}~\bibnamefont
  {Foreman-Mackey}}, \bibinfo {author} {\bibfnamefont {D.~W.}\ \bibnamefont
  {Hogg}}, \bibinfo {author} {\bibfnamefont {D.}~\bibnamefont {Lang}}, \ and\
  \bibinfo {author} {\bibfnamefont {J.}~\bibnamefont {Goodman}},\ }\href
  {\doibase 10.1086/670067} {\bibfield  {journal} {\bibinfo  {journal}
  {Publications of the Astronomical Society of the Pacific}\ }\textbf {\bibinfo
  {volume} {125}},\ \bibinfo {pages} {306} (\bibinfo {year}
  {2013})}\BibitemShut {NoStop}%
\bibitem [{\citenamefont {Ver~Hoef}\ and\ \citenamefont
  {Boveng}(2007)}]{VerHoef2007}%
  \BibitemOpen
  \bibfield  {author} {\bibinfo {author} {\bibfnamefont {J.~M.}\ \bibnamefont
  {Ver~Hoef}}\ and\ \bibinfo {author} {\bibfnamefont {P.~L.}\ \bibnamefont
  {Boveng}},\ }\href {\doibase https://doi.org/10.1890/07-0043.1} {\bibfield
  {journal} {\bibinfo  {journal} {Ecology}\ }\textbf {\bibinfo {volume} {88}},\
  \bibinfo {pages} {2766} (\bibinfo {year} {2007})},\ \Eprint
  {http://arxiv.org/abs/https://esajournals.onlinelibrary.wiley.com/doi/pdf/10.1890/07-0043.1}
  {https://esajournals.onlinelibrary.wiley.com/doi/pdf/10.1890/07-0043.1}
  \BibitemShut {NoStop}%
\bibitem [{\citenamefont {Callahan}\ \emph {et~al.}(2019)\citenamefont
  {Callahan}, \citenamefont {Liu}, \citenamefont {Gonzalez}, \citenamefont
  {Adamek}, \citenamefont {Bowman}, \citenamefont {Broussard}, \citenamefont
  {Clayton}, \citenamefont {Currie}, \citenamefont {Cude-Woods}, \citenamefont
  {Dees}, \citenamefont {Ding}, \citenamefont {Fox}, \citenamefont
  {Geltenbort}, \citenamefont {Hickerson}, \citenamefont {Hoffbauer},
  \citenamefont {Holley}, \citenamefont {Komives}, \citenamefont {MacDonald},
  \citenamefont {Makela}, \citenamefont {Morris}, \citenamefont {Ortiz},
  \citenamefont {Pattie}, \citenamefont {Ramsey}, \citenamefont {Salvat},
  \citenamefont {Saunders}, \citenamefont {Sharapov}, \citenamefont {Sjue},
  \citenamefont {Tang}, \citenamefont {Vanderwerp}, \citenamefont {Vogelaar},
  \citenamefont {Walstrom}, \citenamefont {Wang}, \citenamefont {Weaver},
  \citenamefont {Wei}, \citenamefont {Wexler},\ and\ \citenamefont
  {Young}}]{Callahan2019}%
  \BibitemOpen
  \bibfield  {author} {\bibinfo {author} {\bibfnamefont {N.}~\bibnamefont
  {Callahan}}, \bibinfo {author} {\bibfnamefont {C.-Y.}\ \bibnamefont {Liu}},
  \bibinfo {author} {\bibfnamefont {F.}~\bibnamefont {Gonzalez}}, \bibinfo
  {author} {\bibfnamefont {E.}~\bibnamefont {Adamek}}, \bibinfo {author}
  {\bibfnamefont {J.~D.}\ \bibnamefont {Bowman}}, \bibinfo {author}
  {\bibfnamefont {L.}~\bibnamefont {Broussard}}, \bibinfo {author}
  {\bibfnamefont {S.~M.}\ \bibnamefont {Clayton}}, \bibinfo {author}
  {\bibfnamefont {S.}~\bibnamefont {Currie}}, \bibinfo {author} {\bibfnamefont
  {C.}~\bibnamefont {Cude-Woods}}, \bibinfo {author} {\bibfnamefont {E.~B.}\
  \bibnamefont {Dees}}, \bibinfo {author} {\bibfnamefont {X.}~\bibnamefont
  {Ding}}, \bibinfo {author} {\bibfnamefont {W.}~\bibnamefont {Fox}}, \bibinfo
  {author} {\bibfnamefont {P.}~\bibnamefont {Geltenbort}}, \bibinfo {author}
  {\bibfnamefont {K.~P.}\ \bibnamefont {Hickerson}}, \bibinfo {author}
  {\bibfnamefont {M.~A.}\ \bibnamefont {Hoffbauer}}, \bibinfo {author}
  {\bibfnamefont {A.~T.}\ \bibnamefont {Holley}}, \bibinfo {author}
  {\bibfnamefont {A.}~\bibnamefont {Komives}}, \bibinfo {author} {\bibfnamefont
  {S.~W.~T.}\ \bibnamefont {MacDonald}}, \bibinfo {author} {\bibfnamefont
  {M.}~\bibnamefont {Makela}}, \bibinfo {author} {\bibfnamefont {C.~L.}\
  \bibnamefont {Morris}}, \bibinfo {author} {\bibfnamefont {J.~D.}\
  \bibnamefont {Ortiz}}, \bibinfo {author} {\bibfnamefont {R.~W.}\ \bibnamefont
  {Pattie}}, \bibinfo {author} {\bibfnamefont {J.}~\bibnamefont {Ramsey}},
  \bibinfo {author} {\bibfnamefont {D.~J.}\ \bibnamefont {Salvat}}, \bibinfo
  {author} {\bibfnamefont {A.}~\bibnamefont {Saunders}}, \bibinfo {author}
  {\bibfnamefont {E.~I.}\ \bibnamefont {Sharapov}}, \bibinfo {author}
  {\bibfnamefont {S.~K.~L.}\ \bibnamefont {Sjue}}, \bibinfo {author}
  {\bibfnamefont {Z.}~\bibnamefont {Tang}}, \bibinfo {author} {\bibfnamefont
  {J.}~\bibnamefont {Vanderwerp}}, \bibinfo {author} {\bibfnamefont
  {B.}~\bibnamefont {Vogelaar}}, \bibinfo {author} {\bibfnamefont {P.~L.}\
  \bibnamefont {Walstrom}}, \bibinfo {author} {\bibfnamefont {Z.}~\bibnamefont
  {Wang}}, \bibinfo {author} {\bibfnamefont {H.}~\bibnamefont {Weaver}},
  \bibinfo {author} {\bibfnamefont {W.}~\bibnamefont {Wei}}, \bibinfo {author}
  {\bibfnamefont {J.}~\bibnamefont {Wexler}}, \ and\ \bibinfo {author}
  {\bibfnamefont {A.~R.}\ \bibnamefont {Young}},\ }\href {\doibase
  10.1103/PhysRevC.100.015501} {\bibfield  {journal} {\bibinfo  {journal}
  {Phys. Rev. C}\ }\textbf {\bibinfo {volume} {100}},\ \bibinfo {pages}
  {015501} (\bibinfo {year} {2019})}\BibitemShut {NoStop}%
\bibitem [{\citenamefont {Seestrom}\ \emph {et~al.}(2015)\citenamefont
  {Seestrom}, \citenamefont {Adamek}, \citenamefont {Barlow}, \citenamefont
  {Broussard}, \citenamefont {Callahan}, \citenamefont {Clayton}, \citenamefont
  {Cude-Woods}, \citenamefont {Currie}, \citenamefont {Dees}, \citenamefont
  {Fox}, \citenamefont {Geltenbort}, \citenamefont {Hickerson}, \citenamefont
  {Holley}, \citenamefont {Liu}, \citenamefont {Makela}, \citenamefont
  {Medina}, \citenamefont {Morley}, \citenamefont {Morris}, \citenamefont
  {Ramsey}, \citenamefont {Roberts}, \citenamefont {Salvat}, \citenamefont
  {Saunders}, \citenamefont {Sharapov}, \citenamefont {Sjue}, \citenamefont
  {Slaughter}, \citenamefont {VornDick}, \citenamefont {Walstrom},
  \citenamefont {Wang}, \citenamefont {Womack}, \citenamefont {Young},\ and\
  \citenamefont {Zeck}}]{Seestrom2015}%
  \BibitemOpen
  \bibfield  {author} {\bibinfo {author} {\bibfnamefont {S.~J.}\ \bibnamefont
  {Seestrom}}, \bibinfo {author} {\bibfnamefont {E.~R.}\ \bibnamefont
  {Adamek}}, \bibinfo {author} {\bibfnamefont {D.}~\bibnamefont {Barlow}},
  \bibinfo {author} {\bibfnamefont {L.~J.}\ \bibnamefont {Broussard}}, \bibinfo
  {author} {\bibfnamefont {N.~B.}\ \bibnamefont {Callahan}}, \bibinfo {author}
  {\bibfnamefont {S.~M.}\ \bibnamefont {Clayton}}, \bibinfo {author}
  {\bibfnamefont {C.}~\bibnamefont {Cude-Woods}}, \bibinfo {author}
  {\bibfnamefont {S.}~\bibnamefont {Currie}}, \bibinfo {author} {\bibfnamefont
  {E.~B.}\ \bibnamefont {Dees}}, \bibinfo {author} {\bibfnamefont
  {W.}~\bibnamefont {Fox}}, \bibinfo {author} {\bibfnamefont {P.}~\bibnamefont
  {Geltenbort}}, \bibinfo {author} {\bibfnamefont {K.~P.}\ \bibnamefont
  {Hickerson}}, \bibinfo {author} {\bibfnamefont {A.~T.}\ \bibnamefont
  {Holley}}, \bibinfo {author} {\bibfnamefont {C.-Y.}\ \bibnamefont {Liu}},
  \bibinfo {author} {\bibfnamefont {M.}~\bibnamefont {Makela}}, \bibinfo
  {author} {\bibfnamefont {J.}~\bibnamefont {Medina}}, \bibinfo {author}
  {\bibfnamefont {D.~J.}\ \bibnamefont {Morley}}, \bibinfo {author}
  {\bibfnamefont {C.~L.}\ \bibnamefont {Morris}}, \bibinfo {author}
  {\bibfnamefont {J.}~\bibnamefont {Ramsey}}, \bibinfo {author} {\bibfnamefont
  {A.}~\bibnamefont {Roberts}}, \bibinfo {author} {\bibfnamefont {D.~J.}\
  \bibnamefont {Salvat}}, \bibinfo {author} {\bibfnamefont {A.}~\bibnamefont
  {Saunders}}, \bibinfo {author} {\bibfnamefont {E.~I.}\ \bibnamefont
  {Sharapov}}, \bibinfo {author} {\bibfnamefont {S.~K.~L.}\ \bibnamefont
  {Sjue}}, \bibinfo {author} {\bibfnamefont {B.~A.}\ \bibnamefont {Slaughter}},
  \bibinfo {author} {\bibfnamefont {B.}~\bibnamefont {VornDick}}, \bibinfo
  {author} {\bibfnamefont {P.~L.}\ \bibnamefont {Walstrom}}, \bibinfo {author}
  {\bibfnamefont {Z.}~\bibnamefont {Wang}}, \bibinfo {author} {\bibfnamefont
  {T.~L.}\ \bibnamefont {Womack}}, \bibinfo {author} {\bibfnamefont {A.~R.}\
  \bibnamefont {Young}}, \ and\ \bibinfo {author} {\bibfnamefont {B.~A.}\
  \bibnamefont {Zeck}} (\bibinfo {collaboration} {UCN\ensuremath{\tau}
  Collaboration}),\ }\href {\doibase 10.1103/PhysRevC.92.065501} {\bibfield
  {journal} {\bibinfo  {journal} {Phys. Rev. C}\ }\textbf {\bibinfo {volume}
  {92}},\ \bibinfo {pages} {065501} (\bibinfo {year} {2015})}\BibitemShut
  {NoStop}%
\bibitem [{\citenamefont {Seestrom}\ \emph {et~al.}(2017)\citenamefont
  {Seestrom}, \citenamefont {Adamek}, \citenamefont {Barlow}, \citenamefont
  {Blatnik}, \citenamefont {Broussard}, \citenamefont {Callahan}, \citenamefont
  {Clayton}, \citenamefont {Cude-Woods}, \citenamefont {Currie}, \citenamefont
  {Dees}, \citenamefont {Fox}, \citenamefont {Hoffbauer}, \citenamefont
  {Hickerson}, \citenamefont {Holley}, \citenamefont {Liu}, \citenamefont
  {Makela}, \citenamefont {Medina}, \citenamefont {Morley}, \citenamefont
  {Morris}, \citenamefont {Pattie}, \citenamefont {Ramsey}, \citenamefont
  {Roberts}, \citenamefont {Salvat}, \citenamefont {Saunders}, \citenamefont
  {Sharapov}, \citenamefont {Sjue}, \citenamefont {Slaughter}, \citenamefont
  {Walstrom}, \citenamefont {Wang}, \citenamefont {Wexler}, \citenamefont
  {Womack}, \citenamefont {Young}, \citenamefont {Vanderwerp},\ and\
  \citenamefont {Zeck}}]{Seestrom2017}%
  \BibitemOpen
  \bibfield  {author} {\bibinfo {author} {\bibfnamefont {S.~J.}\ \bibnamefont
  {Seestrom}}, \bibinfo {author} {\bibfnamefont {E.~R.}\ \bibnamefont
  {Adamek}}, \bibinfo {author} {\bibfnamefont {D.}~\bibnamefont {Barlow}},
  \bibinfo {author} {\bibfnamefont {M.}~\bibnamefont {Blatnik}}, \bibinfo
  {author} {\bibfnamefont {L.~J.}\ \bibnamefont {Broussard}}, \bibinfo {author}
  {\bibfnamefont {N.~B.}\ \bibnamefont {Callahan}}, \bibinfo {author}
  {\bibfnamefont {S.~M.}\ \bibnamefont {Clayton}}, \bibinfo {author}
  {\bibfnamefont {C.}~\bibnamefont {Cude-Woods}}, \bibinfo {author}
  {\bibfnamefont {S.}~\bibnamefont {Currie}}, \bibinfo {author} {\bibfnamefont
  {E.~B.}\ \bibnamefont {Dees}}, \bibinfo {author} {\bibfnamefont
  {W.}~\bibnamefont {Fox}}, \bibinfo {author} {\bibfnamefont {M.}~\bibnamefont
  {Hoffbauer}}, \bibinfo {author} {\bibfnamefont {K.~P.}\ \bibnamefont
  {Hickerson}}, \bibinfo {author} {\bibfnamefont {A.~T.}\ \bibnamefont
  {Holley}}, \bibinfo {author} {\bibfnamefont {C.-Y.}\ \bibnamefont {Liu}},
  \bibinfo {author} {\bibfnamefont {M.}~\bibnamefont {Makela}}, \bibinfo
  {author} {\bibfnamefont {J.}~\bibnamefont {Medina}}, \bibinfo {author}
  {\bibfnamefont {D.~J.}\ \bibnamefont {Morley}}, \bibinfo {author}
  {\bibfnamefont {C.~L.}\ \bibnamefont {Morris}}, \bibinfo {author}
  {\bibfnamefont {R.~W.}\ \bibnamefont {Pattie}}, \bibinfo {author}
  {\bibfnamefont {J.}~\bibnamefont {Ramsey}}, \bibinfo {author} {\bibfnamefont
  {A.}~\bibnamefont {Roberts}}, \bibinfo {author} {\bibfnamefont {D.~J.}\
  \bibnamefont {Salvat}}, \bibinfo {author} {\bibfnamefont {A.}~\bibnamefont
  {Saunders}}, \bibinfo {author} {\bibfnamefont {E.~I.}\ \bibnamefont
  {Sharapov}}, \bibinfo {author} {\bibfnamefont {S.~K.~L.}\ \bibnamefont
  {Sjue}}, \bibinfo {author} {\bibfnamefont {B.~A.}\ \bibnamefont {Slaughter}},
  \bibinfo {author} {\bibfnamefont {P.~L.}\ \bibnamefont {Walstrom}}, \bibinfo
  {author} {\bibfnamefont {Z.}~\bibnamefont {Wang}}, \bibinfo {author}
  {\bibfnamefont {J.}~\bibnamefont {Wexler}}, \bibinfo {author} {\bibfnamefont
  {T.~L.}\ \bibnamefont {Womack}}, \bibinfo {author} {\bibfnamefont {A.~R.}\
  \bibnamefont {Young}}, \bibinfo {author} {\bibfnamefont {J.}~\bibnamefont
  {Vanderwerp}}, \ and\ \bibinfo {author} {\bibfnamefont {B.~A.}\ \bibnamefont
  {Zeck}},\ }\href {\doibase 10.1103/PhysRevC.95.015501} {\bibfield  {journal}
  {\bibinfo  {journal} {Phys. Rev. C}\ }\textbf {\bibinfo {volume} {95}},\
  \bibinfo {pages} {015501} (\bibinfo {year} {2017})}\BibitemShut {NoStop}%
\bibitem [{\citenamefont {Albahri}\ \emph {et~al.}(2021)\citenamefont
  {Albahri}, \citenamefont {Anastasi}, \citenamefont {Anisenkov}, \citenamefont
  {Badgley}, \citenamefont {Bae\ss{}ler}, \citenamefont {Bailey}, \citenamefont
  {Baranov}, \citenamefont {Barlas-Yucel}, \citenamefont {Barrett},
  \citenamefont {Basti}, \citenamefont {Bedeschi}, \citenamefont {Berz},
  \citenamefont {Bhattacharya}, \citenamefont {Binney}, \citenamefont {Bloom},
  \citenamefont {Bono}, \citenamefont {Bottalico}, \citenamefont {Bowcock},
  \citenamefont {Cantatore}, \citenamefont {Carey}, \citenamefont {Casey},
  \citenamefont {Cauz}, \citenamefont {Chakraborty}, \citenamefont {Chang},
  \citenamefont {Chapelain}, \citenamefont {Charity}, \citenamefont {Chislett},
  \citenamefont {Choi}, \citenamefont {Chu}, \citenamefont {Chupp},
  \citenamefont {Corrodi}, \citenamefont {Cotrozzi}, \citenamefont {Crnkovic},
  \citenamefont {Dabagov}, \citenamefont {Debevec}, \citenamefont {Di~Falco},
  \citenamefont {Di~Meo}, \citenamefont {Di~Sciascio}, \citenamefont
  {Di~Stefano}, \citenamefont {Driutti}, \citenamefont {Duginov}, \citenamefont
  {Eads}, \citenamefont {Esquivel}, \citenamefont {Farooq}, \citenamefont
  {Fatemi}, \citenamefont {Ferrari}, \citenamefont {Fertl}, \citenamefont
  {Fienberg}, \citenamefont {Fioretti}, \citenamefont {Flay}, \citenamefont
  {Frle\ifmmode~\check{z}\else \v{z}\fi{}}, \citenamefont {Froemming},
  \citenamefont {Fry}, \citenamefont {Gabbanini}, \citenamefont {Galati},
  \citenamefont {Ganguly}, \citenamefont {Garcia}, \citenamefont {George},
  \citenamefont {Gibbons}, \citenamefont {Gioiosa}, \citenamefont {Giovanetti},
  \citenamefont {Girotti}, \citenamefont {Gohn}, \citenamefont {Gorringe},
  \citenamefont {Grange}, \citenamefont {Grant}, \citenamefont {Gray},
  \citenamefont {Haciomeroglu}, \citenamefont {Halewood-Leagas}, \citenamefont
  {Hampai}, \citenamefont {Han}, \citenamefont {Hempstead}, \citenamefont
  {Herrod}, \citenamefont {Hertzog}, \citenamefont {Hesketh}, \citenamefont
  {Hibbert}, \citenamefont {Hodge}, \citenamefont {Holzbauer}, \citenamefont
  {Hong}, \citenamefont {Hong}, \citenamefont {Iacovacci}, \citenamefont
  {Incagli}, \citenamefont {Kammel}, \citenamefont {Kargiantoulakis},
  \citenamefont {Karuza}, \citenamefont {Kaspar}, \citenamefont {Kawall},
  \citenamefont {Kelton}, \citenamefont {Keshavarzi}, \citenamefont {Kessler},
  \citenamefont {Khaw}, \citenamefont {Khechadoorian}, \citenamefont
  {Khomutov}, \citenamefont {Kiburg}, \citenamefont {Kiburg}, \citenamefont
  {Kim}, \citenamefont {Kim}, \citenamefont {King}, \citenamefont {Kinnaird},
  \citenamefont {Kraegeloh}, \citenamefont {Kuchibhotla}, \citenamefont
  {Kuchinskiy}, \citenamefont {Labe}, \citenamefont {LaBounty}, \citenamefont
  {Lancaster}, \citenamefont {Lee}, \citenamefont {Lee}, \citenamefont {Leo},
  \citenamefont {Li}, \citenamefont {Li}, \citenamefont {Li}, \citenamefont
  {Logashenko}, \citenamefont {Lorente~Campos}, \citenamefont {Luc\`a},
  \citenamefont {Lukicov}, \citenamefont {Lusiani}, \citenamefont {Lyon},
  \citenamefont {MacCoy}, \citenamefont {Madrak}, \citenamefont {Makino},
  \citenamefont {Marignetti}, \citenamefont {Mastroianni}, \citenamefont
  {Miller}, \citenamefont {Miozzi}, \citenamefont {Morse}, \citenamefont
  {Mott}, \citenamefont {Nath}, \citenamefont {Nguyen}, \citenamefont
  {Osofsky}, \citenamefont {Park}, \citenamefont {Pauletta}, \citenamefont
  {Piacentino}, \citenamefont {Pilato}, \citenamefont {Pitts}, \citenamefont
  {Plaster}, \citenamefont {Po\ifmmode \check{c}\else
  \v{c}\fi{}ani\ifmmode~\acute{c}\else \'{c}\fi{}}, \citenamefont {Pohlman},
  \citenamefont {Polly}, \citenamefont {Price}, \citenamefont {Quinn},
  \citenamefont {Raha}, \citenamefont {Ramachandran}, \citenamefont {Ramberg},
  \citenamefont {Ritchie}, \citenamefont {Roberts}, \citenamefont {Rubin},
  \citenamefont {Santi}, \citenamefont {Schlesier}, \citenamefont
  {Schreckenberger}, \citenamefont {Semertzidis}, \citenamefont {Shemyakin},
  \citenamefont {Smith}, \citenamefont {Sorbara}, \citenamefont {St\"ockinger},
  \citenamefont {Stapleton}, \citenamefont {Stoughton}, \citenamefont
  {Stratakis}, \citenamefont {Stuttard}, \citenamefont {Swanson}, \citenamefont
  {Sweetmore}, \citenamefont {Sweigart}, \citenamefont {Syphers}, \citenamefont
  {Tarazona}, \citenamefont {Teubner}, \citenamefont {Tewsley-Booth},
  \citenamefont {Thomson}, \citenamefont {Tishchenko}, \citenamefont {Tran},
  \citenamefont {Turner}, \citenamefont {Valetov}, \citenamefont {Vasilkova},
  \citenamefont {Venanzoni}, \citenamefont {Walton}, \citenamefont {Weisskopf},
  \citenamefont {Welty-Rieger}, \citenamefont {Winter}, \citenamefont
  {Wolski},\ and\ \citenamefont {Wu}}]{Albahri2021}%
  \BibitemOpen
  \bibfield  {author} {\bibinfo {author} {\bibfnamefont {T.}~\bibnamefont
  {Albahri}}, \bibinfo {author} {\bibfnamefont {A.}~\bibnamefont {Anastasi}},
  \bibinfo {author} {\bibfnamefont {A.}~\bibnamefont {Anisenkov}}, \bibinfo
  {author} {\bibfnamefont {K.}~\bibnamefont {Badgley}}, \bibinfo {author}
  {\bibfnamefont {S.}~\bibnamefont {Bae\ss{}ler}}, \bibinfo {author}
  {\bibfnamefont {I.}~\bibnamefont {Bailey}}, \bibinfo {author} {\bibfnamefont
  {V.~A.}\ \bibnamefont {Baranov}}, \bibinfo {author} {\bibfnamefont
  {E.}~\bibnamefont {Barlas-Yucel}}, \bibinfo {author} {\bibfnamefont
  {T.}~\bibnamefont {Barrett}}, \bibinfo {author} {\bibfnamefont
  {A.}~\bibnamefont {Basti}}, \bibinfo {author} {\bibfnamefont
  {F.}~\bibnamefont {Bedeschi}}, \bibinfo {author} {\bibfnamefont
  {M.}~\bibnamefont {Berz}}, \bibinfo {author} {\bibfnamefont {M.}~\bibnamefont
  {Bhattacharya}}, \bibinfo {author} {\bibfnamefont {H.~P.}\ \bibnamefont
  {Binney}}, \bibinfo {author} {\bibfnamefont {P.}~\bibnamefont {Bloom}},
  \bibinfo {author} {\bibfnamefont {J.}~\bibnamefont {Bono}}, \bibinfo {author}
  {\bibfnamefont {E.}~\bibnamefont {Bottalico}}, \bibinfo {author}
  {\bibfnamefont {T.}~\bibnamefont {Bowcock}}, \bibinfo {author} {\bibfnamefont
  {G.}~\bibnamefont {Cantatore}}, \bibinfo {author} {\bibfnamefont {R.~M.}\
  \bibnamefont {Carey}}, \bibinfo {author} {\bibfnamefont {B.~C.~K.}\
  \bibnamefont {Casey}}, \bibinfo {author} {\bibfnamefont {D.}~\bibnamefont
  {Cauz}}, \bibinfo {author} {\bibfnamefont {R.}~\bibnamefont {Chakraborty}},
  \bibinfo {author} {\bibfnamefont {S.~P.}\ \bibnamefont {Chang}}, \bibinfo
  {author} {\bibfnamefont {A.}~\bibnamefont {Chapelain}}, \bibinfo {author}
  {\bibfnamefont {S.}~\bibnamefont {Charity}}, \bibinfo {author} {\bibfnamefont
  {R.}~\bibnamefont {Chislett}}, \bibinfo {author} {\bibfnamefont
  {J.}~\bibnamefont {Choi}}, \bibinfo {author} {\bibfnamefont {Z.}~\bibnamefont
  {Chu}}, \bibinfo {author} {\bibfnamefont {T.~E.}\ \bibnamefont {Chupp}},
  \bibinfo {author} {\bibfnamefont {S.}~\bibnamefont {Corrodi}}, \bibinfo
  {author} {\bibfnamefont {L.}~\bibnamefont {Cotrozzi}}, \bibinfo {author}
  {\bibfnamefont {J.~D.}\ \bibnamefont {Crnkovic}}, \bibinfo {author}
  {\bibfnamefont {S.}~\bibnamefont {Dabagov}}, \bibinfo {author} {\bibfnamefont
  {P.~T.}\ \bibnamefont {Debevec}}, \bibinfo {author} {\bibfnamefont
  {S.}~\bibnamefont {Di~Falco}}, \bibinfo {author} {\bibfnamefont
  {P.}~\bibnamefont {Di~Meo}}, \bibinfo {author} {\bibfnamefont
  {G.}~\bibnamefont {Di~Sciascio}}, \bibinfo {author} {\bibfnamefont
  {R.}~\bibnamefont {Di~Stefano}}, \bibinfo {author} {\bibfnamefont
  {A.}~\bibnamefont {Driutti}}, \bibinfo {author} {\bibfnamefont {V.~N.}\
  \bibnamefont {Duginov}}, \bibinfo {author} {\bibfnamefont {M.}~\bibnamefont
  {Eads}}, \bibinfo {author} {\bibfnamefont {J.}~\bibnamefont {Esquivel}},
  \bibinfo {author} {\bibfnamefont {M.}~\bibnamefont {Farooq}}, \bibinfo
  {author} {\bibfnamefont {R.}~\bibnamefont {Fatemi}}, \bibinfo {author}
  {\bibfnamefont {C.}~\bibnamefont {Ferrari}}, \bibinfo {author} {\bibfnamefont
  {M.}~\bibnamefont {Fertl}}, \bibinfo {author} {\bibfnamefont {A.~T.}\
  \bibnamefont {Fienberg}}, \bibinfo {author} {\bibfnamefont {A.}~\bibnamefont
  {Fioretti}}, \bibinfo {author} {\bibfnamefont {D.}~\bibnamefont {Flay}},
  \bibinfo {author} {\bibfnamefont {E.}~\bibnamefont
  {Frle\ifmmode~\check{z}\else \v{z}\fi{}}}, \bibinfo {author} {\bibfnamefont
  {N.~S.}\ \bibnamefont {Froemming}}, \bibinfo {author} {\bibfnamefont
  {J.}~\bibnamefont {Fry}}, \bibinfo {author} {\bibfnamefont {C.}~\bibnamefont
  {Gabbanini}}, \bibinfo {author} {\bibfnamefont {M.~D.}\ \bibnamefont
  {Galati}}, \bibinfo {author} {\bibfnamefont {S.}~\bibnamefont {Ganguly}},
  \bibinfo {author} {\bibfnamefont {A.}~\bibnamefont {Garcia}}, \bibinfo
  {author} {\bibfnamefont {J.}~\bibnamefont {George}}, \bibinfo {author}
  {\bibfnamefont {L.~K.}\ \bibnamefont {Gibbons}}, \bibinfo {author}
  {\bibfnamefont {A.}~\bibnamefont {Gioiosa}}, \bibinfo {author} {\bibfnamefont
  {K.~L.}\ \bibnamefont {Giovanetti}}, \bibinfo {author} {\bibfnamefont
  {P.}~\bibnamefont {Girotti}}, \bibinfo {author} {\bibfnamefont
  {W.}~\bibnamefont {Gohn}}, \bibinfo {author} {\bibfnamefont {T.}~\bibnamefont
  {Gorringe}}, \bibinfo {author} {\bibfnamefont {J.}~\bibnamefont {Grange}},
  \bibinfo {author} {\bibfnamefont {S.}~\bibnamefont {Grant}}, \bibinfo
  {author} {\bibfnamefont {F.}~\bibnamefont {Gray}}, \bibinfo {author}
  {\bibfnamefont {S.}~\bibnamefont {Haciomeroglu}}, \bibinfo {author}
  {\bibfnamefont {T.}~\bibnamefont {Halewood-Leagas}}, \bibinfo {author}
  {\bibfnamefont {D.}~\bibnamefont {Hampai}}, \bibinfo {author} {\bibfnamefont
  {F.}~\bibnamefont {Han}}, \bibinfo {author} {\bibfnamefont {J.}~\bibnamefont
  {Hempstead}}, \bibinfo {author} {\bibfnamefont {A.~T.}\ \bibnamefont
  {Herrod}}, \bibinfo {author} {\bibfnamefont {D.~W.}\ \bibnamefont {Hertzog}},
  \bibinfo {author} {\bibfnamefont {G.}~\bibnamefont {Hesketh}}, \bibinfo
  {author} {\bibfnamefont {A.}~\bibnamefont {Hibbert}}, \bibinfo {author}
  {\bibfnamefont {Z.}~\bibnamefont {Hodge}}, \bibinfo {author} {\bibfnamefont
  {J.~L.}\ \bibnamefont {Holzbauer}}, \bibinfo {author} {\bibfnamefont {K.~W.}\
  \bibnamefont {Hong}}, \bibinfo {author} {\bibfnamefont {R.}~\bibnamefont
  {Hong}}, \bibinfo {author} {\bibfnamefont {M.}~\bibnamefont {Iacovacci}},
  \bibinfo {author} {\bibfnamefont {M.}~\bibnamefont {Incagli}}, \bibinfo
  {author} {\bibfnamefont {P.}~\bibnamefont {Kammel}}, \bibinfo {author}
  {\bibfnamefont {M.}~\bibnamefont {Kargiantoulakis}}, \bibinfo {author}
  {\bibfnamefont {M.}~\bibnamefont {Karuza}}, \bibinfo {author} {\bibfnamefont
  {J.}~\bibnamefont {Kaspar}}, \bibinfo {author} {\bibfnamefont
  {D.}~\bibnamefont {Kawall}}, \bibinfo {author} {\bibfnamefont
  {L.}~\bibnamefont {Kelton}}, \bibinfo {author} {\bibfnamefont
  {A.}~\bibnamefont {Keshavarzi}}, \bibinfo {author} {\bibfnamefont
  {D.}~\bibnamefont {Kessler}}, \bibinfo {author} {\bibfnamefont {K.~S.}\
  \bibnamefont {Khaw}}, \bibinfo {author} {\bibfnamefont {Z.}~\bibnamefont
  {Khechadoorian}}, \bibinfo {author} {\bibfnamefont {N.~V.}\ \bibnamefont
  {Khomutov}}, \bibinfo {author} {\bibfnamefont {B.}~\bibnamefont {Kiburg}},
  \bibinfo {author} {\bibfnamefont {M.}~\bibnamefont {Kiburg}}, \bibinfo
  {author} {\bibfnamefont {O.}~\bibnamefont {Kim}}, \bibinfo {author}
  {\bibfnamefont {Y.~I.}\ \bibnamefont {Kim}}, \bibinfo {author} {\bibfnamefont
  {B.}~\bibnamefont {King}}, \bibinfo {author} {\bibfnamefont {N.}~\bibnamefont
  {Kinnaird}}, \bibinfo {author} {\bibfnamefont {E.}~\bibnamefont {Kraegeloh}},
  \bibinfo {author} {\bibfnamefont {A.}~\bibnamefont {Kuchibhotla}}, \bibinfo
  {author} {\bibfnamefont {N.~A.}\ \bibnamefont {Kuchinskiy}}, \bibinfo
  {author} {\bibfnamefont {K.~R.}\ \bibnamefont {Labe}}, \bibinfo {author}
  {\bibfnamefont {J.}~\bibnamefont {LaBounty}}, \bibinfo {author}
  {\bibfnamefont {M.}~\bibnamefont {Lancaster}}, \bibinfo {author}
  {\bibfnamefont {M.~J.}\ \bibnamefont {Lee}}, \bibinfo {author} {\bibfnamefont
  {S.}~\bibnamefont {Lee}}, \bibinfo {author} {\bibfnamefont {S.}~\bibnamefont
  {Leo}}, \bibinfo {author} {\bibfnamefont {B.}~\bibnamefont {Li}}, \bibinfo
  {author} {\bibfnamefont {D.}~\bibnamefont {Li}}, \bibinfo {author}
  {\bibfnamefont {L.}~\bibnamefont {Li}}, \bibinfo {author} {\bibfnamefont
  {I.}~\bibnamefont {Logashenko}}, \bibinfo {author} {\bibfnamefont
  {A.}~\bibnamefont {Lorente~Campos}}, \bibinfo {author} {\bibfnamefont
  {A.}~\bibnamefont {Luc\`a}}, \bibinfo {author} {\bibfnamefont
  {G.}~\bibnamefont {Lukicov}}, \bibinfo {author} {\bibfnamefont
  {A.}~\bibnamefont {Lusiani}}, \bibinfo {author} {\bibfnamefont {A.~L.}\
  \bibnamefont {Lyon}}, \bibinfo {author} {\bibfnamefont {B.}~\bibnamefont
  {MacCoy}}, \bibinfo {author} {\bibfnamefont {R.}~\bibnamefont {Madrak}},
  \bibinfo {author} {\bibfnamefont {K.}~\bibnamefont {Makino}}, \bibinfo
  {author} {\bibfnamefont {F.}~\bibnamefont {Marignetti}}, \bibinfo {author}
  {\bibfnamefont {S.}~\bibnamefont {Mastroianni}}, \bibinfo {author}
  {\bibfnamefont {J.~P.}\ \bibnamefont {Miller}}, \bibinfo {author}
  {\bibfnamefont {S.}~\bibnamefont {Miozzi}}, \bibinfo {author} {\bibfnamefont
  {W.~M.}\ \bibnamefont {Morse}}, \bibinfo {author} {\bibfnamefont
  {J.}~\bibnamefont {Mott}}, \bibinfo {author} {\bibfnamefont {A.}~\bibnamefont
  {Nath}}, \bibinfo {author} {\bibfnamefont {H.}~\bibnamefont {Nguyen}},
  \bibinfo {author} {\bibfnamefont {R.}~\bibnamefont {Osofsky}}, \bibinfo
  {author} {\bibfnamefont {S.}~\bibnamefont {Park}}, \bibinfo {author}
  {\bibfnamefont {G.}~\bibnamefont {Pauletta}}, \bibinfo {author}
  {\bibfnamefont {G.~M.}\ \bibnamefont {Piacentino}}, \bibinfo {author}
  {\bibfnamefont {R.~N.}\ \bibnamefont {Pilato}}, \bibinfo {author}
  {\bibfnamefont {K.~T.}\ \bibnamefont {Pitts}}, \bibinfo {author}
  {\bibfnamefont {B.}~\bibnamefont {Plaster}}, \bibinfo {author} {\bibfnamefont
  {D.}~\bibnamefont {Po\ifmmode \check{c}\else
  \v{c}\fi{}ani\ifmmode~\acute{c}\else \'{c}\fi{}}}, \bibinfo {author}
  {\bibfnamefont {N.}~\bibnamefont {Pohlman}}, \bibinfo {author} {\bibfnamefont
  {C.~C.}\ \bibnamefont {Polly}}, \bibinfo {author} {\bibfnamefont
  {J.}~\bibnamefont {Price}}, \bibinfo {author} {\bibfnamefont
  {B.}~\bibnamefont {Quinn}}, \bibinfo {author} {\bibfnamefont
  {N.}~\bibnamefont {Raha}}, \bibinfo {author} {\bibfnamefont {S.}~\bibnamefont
  {Ramachandran}}, \bibinfo {author} {\bibfnamefont {E.}~\bibnamefont
  {Ramberg}}, \bibinfo {author} {\bibfnamefont {J.~L.}\ \bibnamefont
  {Ritchie}}, \bibinfo {author} {\bibfnamefont {B.~L.}\ \bibnamefont
  {Roberts}}, \bibinfo {author} {\bibfnamefont {D.~L.}\ \bibnamefont {Rubin}},
  \bibinfo {author} {\bibfnamefont {L.}~\bibnamefont {Santi}}, \bibinfo
  {author} {\bibfnamefont {C.}~\bibnamefont {Schlesier}}, \bibinfo {author}
  {\bibfnamefont {A.}~\bibnamefont {Schreckenberger}}, \bibinfo {author}
  {\bibfnamefont {Y.~K.}\ \bibnamefont {Semertzidis}}, \bibinfo {author}
  {\bibfnamefont {D.}~\bibnamefont {Shemyakin}}, \bibinfo {author}
  {\bibfnamefont {M.~W.}\ \bibnamefont {Smith}}, \bibinfo {author}
  {\bibfnamefont {M.}~\bibnamefont {Sorbara}}, \bibinfo {author} {\bibfnamefont
  {D.}~\bibnamefont {St\"ockinger}}, \bibinfo {author} {\bibfnamefont
  {J.}~\bibnamefont {Stapleton}}, \bibinfo {author} {\bibfnamefont
  {C.}~\bibnamefont {Stoughton}}, \bibinfo {author} {\bibfnamefont
  {D.}~\bibnamefont {Stratakis}}, \bibinfo {author} {\bibfnamefont
  {T.}~\bibnamefont {Stuttard}}, \bibinfo {author} {\bibfnamefont {H.~E.}\
  \bibnamefont {Swanson}}, \bibinfo {author} {\bibfnamefont {G.}~\bibnamefont
  {Sweetmore}}, \bibinfo {author} {\bibfnamefont {D.~A.}\ \bibnamefont
  {Sweigart}}, \bibinfo {author} {\bibfnamefont {M.~J.}\ \bibnamefont
  {Syphers}}, \bibinfo {author} {\bibfnamefont {D.~A.}\ \bibnamefont
  {Tarazona}}, \bibinfo {author} {\bibfnamefont {T.}~\bibnamefont {Teubner}},
  \bibinfo {author} {\bibfnamefont {A.~E.}\ \bibnamefont {Tewsley-Booth}},
  \bibinfo {author} {\bibfnamefont {K.}~\bibnamefont {Thomson}}, \bibinfo
  {author} {\bibfnamefont {V.}~\bibnamefont {Tishchenko}}, \bibinfo {author}
  {\bibfnamefont {N.~H.}\ \bibnamefont {Tran}}, \bibinfo {author}
  {\bibfnamefont {W.}~\bibnamefont {Turner}}, \bibinfo {author} {\bibfnamefont
  {E.}~\bibnamefont {Valetov}}, \bibinfo {author} {\bibfnamefont
  {D.}~\bibnamefont {Vasilkova}}, \bibinfo {author} {\bibfnamefont
  {G.}~\bibnamefont {Venanzoni}}, \bibinfo {author} {\bibfnamefont
  {T.}~\bibnamefont {Walton}}, \bibinfo {author} {\bibfnamefont
  {A.}~\bibnamefont {Weisskopf}}, \bibinfo {author} {\bibfnamefont
  {L.}~\bibnamefont {Welty-Rieger}}, \bibinfo {author} {\bibfnamefont
  {P.}~\bibnamefont {Winter}}, \bibinfo {author} {\bibfnamefont
  {A.}~\bibnamefont {Wolski}}, \ and\ \bibinfo {author} {\bibfnamefont
  {W.}~\bibnamefont {Wu}} (\bibinfo {collaboration} {Muon $g\ensuremath{-}2$
  Collaboration}),\ }\href {\doibase 10.1103/PhysRevD.103.072002} {\bibfield
  {journal} {\bibinfo  {journal} {Phys. Rev. D}\ }\textbf {\bibinfo {volume}
  {103}},\ \bibinfo {pages} {072002} (\bibinfo {year} {2021})}\BibitemShut
  {NoStop}%
\end{thebibliography}%

\end{document}